\documentclass{article}
\pdfoutput=1
\usepackage{jcappub}
\usepackage{amsmath}
\usepackage{amssymb}
\usepackage{hyperref}
\usepackage{array,multirow}

\definecolor{darkred}{RGB}{175,0,0}

\title{Constraining extended cosmologies with GW$\times$LSS cross-correlations}

\author[a]{M. Bosi,}
\emailAdd{michele.bosi@studenti.unipd.it}

\author[b]{N. Bellomo,}
\emailAdd{nicola.bellomo@austin.utexas.edu}

\author[a,c,d,e]{A. Raccanelli}
\emailAdd{alvise.raccanelli.1@unipd.it}
\affiliation[a]{Dipartimento di Fisica e Astronomia G. Galilei, Università degli Studi di Padova, via Marzolo 8, I-35131 Padova, Italy.}
\affiliation[b]{Texas Center for Cosmology and Astroparticle Physics, Weinberg Institute, Department of Physics, The University of Texas at Austin, Austin, TX 78712, USA.} 
\affiliation[c]{INFN, Sezione di Padova, via F. Marzolo 8, I-35131 Padova, Italy.}
\affiliation[d]{INAF Padova}
\affiliation[e]{Theoretical Physics Department, CERN, 1 Esplanade des Particules, 1211 Geneva 23,
Switzerland}

\abstract{
The rapid development of gravitational wave astronomy provides the unique opportunity of exploring the dynamics of the Universe using clustering properties of coalescing binary black hole mergers. 
Gravitational wave data, along with information coming from future galaxy surveys, have the potential of shedding light about many open questions in Cosmology, including those regarding the nature of dark matter and dark energy. 
In this work we explore which combination of gravitational wave and galaxy survey datasets are able to provide the best constraints both on modified gravity theories and on the nature of the very same binary black hole events. 
In particular, by using the public Boltzmann code \texttt{Multi\_CLASS}, we compare cosmological constraints on popular~$\Lambda$CDM extensions coming from gravitational waves alone and in conjunction with either deep and localized or wide and shallow galaxy surveys. 
We show that constraints on extensions of General Relativity will be at the same level of existing limits from gravitational waves alone or one order of magnitude better when galaxy surveys are included.
Furthermore, cross-correlating both kind of galaxy survey with gravitational waves datasets will allow to confidently rule in or out primordial black holes as dark matter candidate in the majority of the allowed parameter space.
}

\begin{document}

\hfill{\small UTWI-20-2023}

\maketitle

\section{Introduction}

The~$\Lambda$CDM, i.e., the cosmological Standard Model, passed numerous tests and provided a convincing explanation for the physics behind many cosmological observables, both from the early~\cite{aghanim:planckcosmoparameters} and late Universe~\cite{alam:ebosscosmoparameters, abbott:descosmoparameters, hildebrandt:kidsviking}. 
However, some tensions are still present~\cite{abdalla:cosmologicaltensions}, and a deeper investigation might in fact reveal hints of so-called new physics.

One potential avenue to investigate these tensions is to include in the analysis new and independent datasets. 
In this respect, gravitational waves (GWs) present the unprecedented opportunity to explore the properties and evolution of the Universe in a completely new fashion, complementary to that of more traditional cosmological probes. 
Thanks to the extraordinary effort of the GW community, we have already detected almost a hundred GW events~\cite{abbott:gwtc1, abbott:gwtc2, abbott:gwtc3}, and the number is expected to rapidly grow for every future LIGO-Virgo-KAGRA Collaboration run. 
Moreover, once next generation GW observatories like Einstein Telescope~\cite{maggiore:einsteintelescopewhitepaper} (ET) and Cosmic Explorer~\cite{evans:cosmicexplorerwhitepaper} (CE) will start operating during the next decade, the number of detected events per year will be of order~$\mathcal{O}(10^4-10^5)$, allowing for the unprecedented opportunity to map the entire sky using GWs.

Similarly to electromagnetic radiation, GWs carry information about the sources generating them, the environment where these sources reside and about the Universe the gravitational radiation travelled across. 
In this work we show how combining GW and Large Scale Structure (LSS) datasets we can extract both astrophysical and cosmological information, shedding light both on the nature of dark matter and gravity. 

In recent years, Primordial Black Holes (PBHs) emerged as a possible candidate to make up a non-negligible fraction of the dark matter.
PBHs are (still hypothetical) black holes that -- in the most standard scenario -- formed in the early Universe before big bang nucleosynthesis, deep in the radiation-dominated era~\cite{zeldovich:pbhformation, hawking:pbhformation, carr:pbhformation, chapline:pbhformation}, and since they are not generated through a stellar process, they can span a wide range of masses.
In particular, if PBH masses are larger than~$M_\mathrm{PBH} \gtrsim 10^{-18}\ M_\odot$, their present day abundance can account for a fraction or even the totality of dark matter~\cite{sasaki:pbhconstraintsreview, carr:pbhconstraintsreview}. 
Moreover, similarly to astrophysical BHs, PBHs can form binaries and be detected by present and future ground-based GW observatories if they have masses of order~$\mathcal{O}(1-10^2)\ M_\odot$.
However, PBH binaries (and their relative GW signal) are expected to trace LSS differently from binaries of astrophysical BHs, providing a potential way to infer the nature of the binary constituents~\cite{raccanelli:gwxlss, scelfo:gwxlssI, scelfo:gwxlssII}.

On the other hand, assuming that GWs are generated exclusively by astrophysical sources, we can use their anisotropic distribution to test General Relativity and its extensions~\cite{camera:gwpropagation, namikawa:gwpropagation, mukherjee:gwpropagationI, mukherjee:gwpropagationII, mukherjee:gwpropagationIII, mukherjee:gwpropagationIV, mukherjee:gwpropagationV, garoffolo:gwpropagationI, garoffolo:gwpropagationII, tasinato:gwpropagation, raccanelli:radiosurveysII, scelfo:gwxlssII, scelfo:gwxlssIII, balaudo:gwxlss, bertacca:perturbedluminositydistance, begnoni:perturbedluminositydistance}.
Despite observational evidence seems to prefer~$\Lambda$CDM over alternative models~\cite{ade:planckmodifiedgravity}, the Modified Gravity (MG) community has proposed several theoretical models that extend General Relativity in an effort to explain the accelerated expansion of the Universe~\cite{durrer:modifiedgravityreview, clifton:modifiedgravityreview, ishak:modifiedgravityreview}. 
Present and future LSS surveys plan to further tighten constraints on extension of General Relativity~\cite{norris:emu, abell:lsstsciencebook, aghamousa:desi, laureijs:euclidsciencebook, dore:spherexwhitepaperI, dore:spherexwhitepaperII, dore:spherexwhitepaperIII, spergel:wfirstwhitepaper, maartens:ska, bacon:skaredbook, wang:atlaswhitepaperI,wang:atlaswhitepaperII}, hence it is extremely timely to investigate how new probes can join this effort.

In this work we study if galaxy surveys, deep and localized (DL), or shallow and wide (SW), are suited for extracting cosmological information in combination with GW datasets. 
On the GW side, we consider binaries made of either astrophysical or primordial BHs detected by second and third generation detectors. 
The GW events are modeled using the external modules of \texttt{CLASS\_GWB}~\cite{bellomo:classgwb}, while the galaxy and GW clustering signal is computed by extending the~\texttt{Multi\_CLASS}~\cite{bellomo:multiclass, bernal:multiclass} Boltzmann code\footnote{The code is publicly available at \url{https://github.com/nbellomo/Multi\textunderscore CLASS}.} to account for modified gravity models. 

We show that both kinds of survey we consider will be effective in constraining different extensions of the~$\Lambda$CDM model. 
Regarding the modified gravity/dark energy sector, we show that common extensions of General Relativity like scalar-tensor theories or nDGP models can be constrained by GW alone using only linear scales, obtaining limits comparable to current ones, in a completely independent way and without relying on non-linearity modeling.
When cross-correlating with galaxy surveys, constraints improve by almost an order of magnitude, in similar ways for the dark energy models we considered.
On the other hand, when investigating potential dark matter candidates, we find that the cross-correlation of galaxies and GW can definitively rule in or out with high degree of confidence solar mass PBHs as the main component of dark matter.

This article is structured as follows: in section~\ref{sec:formalism_and_methodology} we present our formalism and the statistical tools we use to forecast the cosmological constraints. 
In section~\ref{sec:tracers} we characterize the tracer we are considering, i.e., galaxies and gravitational waves. 
In section~\ref{sec:results} we quantify which kind of survey gives the best constraints. Finally we conclude in section~\ref{sec:conclusions}. 
Appendices provide additional information regarding the number count fluctuation in General Relativity (\ref{app:numbercountfluctuation_transferfunction}), modifications implemented in~\texttt{Multi\_CLASS} (\ref{app:MG_Multi_CLASS}), second and third generation GW detector network specifications (\ref{app:detector_networks}), PBH binary formation channels (\ref{app:lpbh_binary_formation}), optimistic versions of Modified Gravity parameters (\ref{app:optimistic_mg_constraints}) and astrophysical-primordial BHs second generation detected events (\ref{app:mixed_BBH_2G}).


\section{Formalism and methodology}
\label{sec:formalism_and_methodology}

In this section we briefly summarize the most important aspects of the galaxy and gravitational wave clustering, along with the basic tools we use to assess the constraining power of future galaxy surveys and GW observatory datasets. 
The interested reader can find a broader and richer discussion of these topics in refs.~\cite{bellomo:multiclass, bertacca:gwbprojectioneffects}.


\subsection{Clustering statistics}
\label{subsec:clustering_statistics}

Anisotropies of the Universe LSS contain a wealth of information concerning the origin and the growth of cosmological perturbations. 
In this work we are interested in the fluctuation of the number density of objects~$\Delta^X(z,\hat{\mathbf{n}})$ at redshift~$z$ in the direction~$\hat{\mathbf{n}}$~\cite{bonvin:numbercountfluctuation, challinor:numbercountfluctuation, jeong:numbercountfluctuation}, where~$X$ labels the tracer, i.e., the objects we observe (galaxies or GWs). 
Given the spherical symmetry of the sky, it is convenient to expand the number density fluctuation in spherical harmonics as
\begin{equation}
    \Delta^X(z,\hat{\mathbf{n}})=\sum_{\ell m} a^{X,z}_{\ell m} Y_{\ell m}(\hat{\mathbf{n}}),
\end{equation}
where the~$a^{X,z}_{\ell m}$ are the spherical harmonic coefficients and $Y_{\ell m}$ are the spherical harmonics. 
The two-point statistics in harmonic space, i.e., the angular power spectrum~$C_\ell$, is given by~\cite{raccanelli:crosscorrelation, pullen:crosscorrelation}
\begin{equation}
    \left\langle a^{X,z_i}_{\ell m}a^{Y,z_j *}_{\ell' m'}\right\rangle = \delta^K_{\ell\ell'}\delta^K_{mm'}C_\ell^{XY}(z_i,z_j),
\end{equation}
where~$^*$ indicate the complex conjugate and~$\delta^K$ is the Kronecker delta. 
The angular power spectrum is defined as~\cite{blas:class, didio:classgal}
\begin{equation}
    C_\ell^{XY}(z_i,z_j) = 4 \pi \int\frac{dk}{k} \mathcal{P}(k) \Delta^{X, z_i}_\ell(k)\Delta^{Y, z_j}_\ell(k),
\end{equation}
where~$\mathcal{P}(k)=k^3P(k)/2\pi^2$ is the almost scale-invariant primordial power spectrum. 
The number count fluctuation is
\begin{equation}
    \Delta^{X, z_i}_\ell(k) = \int^{\infty}_0 dz\frac{dN_\mathrm{X}}{dz}W(z,z_i,\Delta z_i)\Delta^{\mathrm{X}}_\ell(k,z),
\end{equation}
where~$dN_\mathrm{X}/dz$ is the redshift distribution of the tracer~$X$ and~$W(z,z_i,\Delta z_i)$ is a window function (normalized to unity) centered at redshift~$z_i$ with half-width~$\Delta z$. 
The number density fluctuation transfer functions~$\Delta^{X}_\ell(k,z)$ is given by the sum of different contribution commonly called density, velocity, lensing and gravity effects~\cite{bonvin:numbercountfluctuation, challinor:numbercountfluctuation, jeong:numbercountfluctuation}:
\begin{equation}
    \Delta^{X}_\ell(k,z) = \Delta^{X,\mathrm{den}}_\ell(k,z) + \Delta^{X,\mathrm{vel}}_\ell(k,z) + \Delta^{X,\mathrm{len}}_\ell(k,z) + \Delta^{X,\mathrm{gr}}_\ell(k,z),
\label{eq:numbercountfluctuation_projectioneffects}
\end{equation}
and the explicit form of these terms is reported in appendix~\ref{app:numbercountfluctuation_transferfunction}. 
The angular power spectra are computed using \texttt{Multi\_CLASS}~\cite{bellomo:multiclass, bernal:multiclass}, a public extension of the Boltzmann code~\texttt{CLASS}~\cite{blas:class} that allows to compute the angular power spectra for different combinations of tracers including all the projection effects included in equation~\eqref{eq:numbercountfluctuation_projectioneffects}. 

Any tracer population is characterized by four functions of redshift and, possibly, scale: the number density redshift distribution~$d^2N_X/dzd\Omega$; the clustering bias~$b_X$, which connects the tracer overdensity to the matter overdensity by~$\delta_X = b_X\delta_\mathrm{m}$; the magnification bias~\cite{matsubara:magnification_bias, hui:magnification_bias, liu:magnification_bias, montanari:magnification_bias}
\begin{equation}
    s_X = -\frac{2}{5} \left. \frac{d\log_{10}\frac{d^2N_X(z, L>L_{\mathrm{lim}})}{dzd\Omega}}{d\log_{10}L} \right|_{L_{\mathrm{lim}}},
\end{equation}
that describes how a flux-limited observable is affected by cosmic lensing effects; and the evolution bias~\cite{challinor:numbercountfluctuation, jeong:numbercountfluctuation, bertacca:evolutionbias}
\begin{equation}
    f_X^\mathrm{evo} = \frac{d\log\left(a^3\frac{d^2N_X}{dzd\Omega}\right)}{d\log a},
\end{equation}
which describes the formation rate of new tracers. 
Moreover, for the purpose of this analysis, we introduce a new parameter called effective bias
\begin{equation}
    b^\mathrm{eff}_X = \int_{z_\mathrm{min}}^{z_\mathrm{max}} dz\ b_X \frac{d^2N_X}{dzd\Omega} \Bigg/ \int_{z_\mathrm{min}}^{z_\mathrm{max}} dz \frac{d^2N_X}{dzd\Omega},
\label{eq:eff_bias}
\end{equation}
where~$z_\mathrm{min}$ and~$z_\mathrm{max}$ are the minimum and maximum redshift of the tracer~$X$ survey, respectively.

Finally, real measurements are affected by multiple sources of noise, which we describe via the noise angular power spectrum~$N^{XY}_\ell (z_i, z_j)$.
In other words, the observed angular power spectrum~$\tilde{C}_\ell$ has the form~$\tilde{C}^{XY}_\ell(z_i,z_j) = C_\ell^{XY}(z_i,z_j) + N^{XY}_\ell(z_i,z_j)$.
The first source of noise we consider is the shot noise due to the discrete nature of the tracer, both galaxies and GWs, which is characterized by a white noise of the form
\begin{equation}
    N^{XY}_\ell(z_i,z_j) = \frac{\delta^K_{XY}\delta^{K}_{ij}}{dN_X(z_i)/d\Omega}.
\label{eq:shot_noise}
\end{equation}
Errors due to poor galaxy localisation, both in terms of angular position and rdshift determination, are expected to be subdominant in our analysis.

However, contrarily to galaxies, GWs suffer from a limited spatial resolution due to instrumental noise.
We take an approach similar to that used in weak lensing analysis, see e.g., refs.~\cite{padmanabhan:weaklensingcalibration, ma:weaklensingcalibration, huterer:weaklensingcalibration}, to treat the error due to poor determination of the GW redshift.
First we define a probability~$p_\mathrm{obs}$ for a GW emitted at redshift~$z_\mathrm{GW}$ to be observed at redshift~$z_\mathrm{obs}$ due to instrumental noise, where, for simplicity, we assume
\begin{equation}
    p_\mathrm{obs}(z_\mathrm{GW},z_\mathrm{obs}) = \frac{1}{\sqrt{2\pi\sigma^2_z}} e^{-(z_\mathrm{GW}-z_\mathrm{obs})^2/2\sigma^2_z}.
\end{equation}
In the case of an infinitely precise instrument the dispersion~$\sigma_z$ tends to zero and the probability~$p_\mathrm{obs}$ becomes a Dirac delta centered at the true value~$z_\mathrm{GW}$, as we naively expect to be the case.
Then we introduce an observed angular number of objects~$dN^\mathrm{obs}_\mathrm{GW}/d\Omega$ and we use this quantity in the shot noise angular power spectrum instead of the true angular number of objects, as done in equation~\eqref{eq:shot_noise}.
More explicitly, for each redshift bin we have
\begin{equation}
    \begin{aligned}
            \frac{dN^\mathrm{obs}_\mathrm{GW}}{d\Omega} &= \int^{z_\mathrm{max}}_{z_\mathrm{min}} dz_\mathrm{obs} \frac{d^2N^\mathrm{obs}_\mathrm{GW}}{dzd\Omega}(z_\mathrm{obs}) = \int^{z_\mathrm{max}}_{z_\mathrm{min}} dz_\mathrm{obs} \int_0^\infty dz_\mathrm{GW} \frac{d^2N_\mathrm{GW}}{dzd\Omega}(z_\mathrm{GW}) p_\mathrm{obs}(z_\mathrm{GW},z_\mathrm{obs}) \\
            &= \int_0^\infty dz_\mathrm{GW}  \frac{d^2N_\mathrm{GW}}{dzd\Omega}(z_\mathrm{GW}) \frac{1}{2} \left[ \mathrm{erf} \left(\frac{z_\mathrm{GW}-z_\mathrm{min}}{\sqrt{2}\sigma_z}\right) - \mathrm{erf} \left(\frac{z_\mathrm{GW}-z_\mathrm{max}}{\sqrt{2}\sigma_z}\right) \right],
    \end{aligned}
\end{equation}
where, also in this case, in the limit of no instrumental error~$\sigma_z\to 0$ the function in square brackets becomes~$2\Theta_H(z_\mathrm{GW}-z_\mathrm{min}) \Theta_H(z_\mathrm{max}-z_\mathrm{GW})$ and we recover the expected form of the shot noise.
On the other hand, we describe the effect of poor angular resolution via a Gaussian beam enhancing factor of the noise angular power spectrum, as suggested in ref.~\cite{calore:gwxlss}.
Therefore the total GW angular power spectrum~$\tilde{C}_\ell$ has the form
\begin{equation}
    \begin{aligned}
        \tilde{C}^\mathrm{GWGW}_\ell(z_i,z_j) &= C_\ell^\mathrm{GWGW}(z_i,z_j) + N^\mathrm{GWGW}_\ell(z_i,z_j) \\
        &= C_\ell^\mathrm{GWGW}(z_i,z_j) + \frac{\delta^{K}_{ij}}{dN^\mathrm{obs}_\mathrm{GW}(z_i)/d\Omega} \mathrm{exp}\left[\frac{\ell(\ell+1)\left(\theta^\mathrm{avg}_\mathrm{res}\right)^2}{8\log 2}\right],
    \end{aligned}
\label{eq:gw_total_noise}
\end{equation}
where~$\theta^\mathrm{avg}_\mathrm{res}$ is the average resolution of a GW detector network and~``$\log$'' refers to the natural logarithm. 
The maximum multipole~$\ell_\mathrm{max} =  180^\circ/\theta^\mathrm{max}_\mathrm{res}$ we include in our analysis is determined by the maximum resolution of a GW detector network~$\theta^\mathrm{max}_\mathrm{res}$.


\subsection{Fisher matrix analysis}
\label{subsec:fisher_matrix}

Despite its simplicity, a Fisher matrix analysis is able to quickly forecast the constraining power of future experiments. 
In particular, by estimating the curvature of the log-likelihood around its maximum, the Fisher analysis returns the best error an experiment can achieve, i.e., the Cram\'er-Rao bound~\cite{fisher:fishermatrix, bunn:fishermatrix, vogeley:fishermatrix, tegmark:fishermatrix}. 
Given a pair of cosmological parameters~$\{\theta_\alpha,\theta_\beta\}$, their respective Fisher matrix element reads as
\begin{equation}
    F_{\alpha\beta} = \left\langle - \frac{\partial^2\log\mathcal{L}}{\partial\theta_\alpha\partial\theta_\beta} \right\rangle = f_{\mathrm{sky}} \sum_{\ell}\frac{2\ell+1}{2}\mathrm{Tr}\left[\mathcal{C}^{-1}_{\ell}\left(\partial_{\alpha}\mathcal{C}_\ell\right)\mathcal{C}^{-1}_{\ell}\left(\partial_{\beta}\mathcal{C}_\ell\right)\right],
\label{eq:fisher_matrix}
\end{equation}
where to compute the RHS we assume a Gaussian likelihood~$\mathcal{L}$ for the spherical harmonic coefficients, see, e.g., ref.~\cite{bellomo:multiclass}. 
The parameter~$f_\mathrm{sky}$ represents the observed fraction of sky in common between the different (galaxy or GW) surveys considered in the forecast. 
Assuming the tracers are binned into~$n$ redshift bins~$z_1,...,z_n$, the (symmetric) covariance matrix~$\mathcal{C}_\ell$ reads as
\begin{equation}
    \mathcal{C}_{\ell} =
    \begin{pmatrix}
    \tilde C_{\ell}^{\mathrm{g}\mathrm{g}}(z_1,z_1) & \dots &  \tilde C_{\ell}^{\mathrm{g}\mathrm{g}}(z_1,z_n) & \tilde C_{\ell}^{\mathrm{g}\mathrm{GW}}(z_1,z_1) & \dots & \tilde C_{\ell}^{\mathrm{g}\mathrm{GW}}(z_1,z_n) \\
    & \ddots & \vdots & \vdots &  & \vdots \\
    &  & \tilde C_{\ell}^{\mathrm{g}\mathrm{g}}(z_n,z_n) & \tilde C_{\ell}^{\mathrm{g}\mathrm{GW}}(z_n,z_1) & \dots & \tilde C_{\ell}^{\mathrm{g}\mathrm{GW}}(z_n,z_n) \\
    &  &  & \tilde C_{\ell}^{\mathrm{GW}\mathrm{GW}}(z_1,z_1)& \dots &  \tilde C_{\ell}^{\mathrm{GW}\mathrm{GW}}(z_1,z_n) \\
    &  &  &  & \ddots & \vdots \\      
    &  &  &  &  & \tilde C_{\ell}^{\mathrm{GW}\mathrm{GW}}(z_n,z_n)
    \end{pmatrix}.
\end{equation}
The matrix~$\partial_\alpha \mathcal{C}_\ell \equiv \partial\mathcal{C}_\ell/\partial\theta_\alpha$ contains derivatives of the elements of the covariance matrix with respect a single cosmological parameter keeping all the others fixed. 

Errors on cosmological parameters are obtained from the error covariance matrix~$\Sigma=F^{-1}$, in particular the marginalized errors are given by~$\sigma_{\theta_\alpha} = \sqrt{\Sigma_{\alpha\alpha}}$. 
Moreover, using the marginalized errors we can perform a sort of null hypothesis testing comparing two models, the ``fiducial'' and the ``alternative'', each one characterized by the value of a parameter, in our case the GW effective bias.
By computing a Signal-to-Noise ratio (SNR) of the form
\begin{equation}
    \mathrm{SNR}^2 = \frac{\left(b_\mathrm{GW,alt}^\mathrm{eff} - b_\mathrm{GW,fid}^\mathrm{eff}\right)^2}{\sigma^2_{b^\mathrm{eff}_\mathrm{GW,fid}}},
\label{eq:SNR}
\end{equation}
we can assess whether the alternative model, labelled by ``alt'', is statistically different from the fiducial one, labelled by ``fid''.
For values of~$\mathrm{SNR} \lesssim 1$ we consider the two models as statistically indistinguishable.


\subsection{Cosmological model}
\label{subsec:cosmological_model}

In this work we investigate both the nature of dark matter and that of dark energy using GWs, however, given the peculiarity of each scenario, we have to assume different cosmological fiducial models for the two scenarios. 
In the case where we study the origin of coalescing BHs, we assume a standard~$\Lambda$CDM model, in which the gravitational section is described by General Relativity and a cosmological constant is responsible for the accelerated expansion. 
On the other hand, in the second case, where we investigate the nature of dark energy, we allow for deviations from General Relativity. 
Many different theories have been proposed to extend General Relativity by introducing some dynamical mechanism that explains the late time acceleration~\cite{clifton:modifiedgravityreview, ishak:modifiedgravityreview}, however in this article we focus only on two classes, scalar-tensor and extra-dimension theories. 

The most common class of Modified Gravity models are scalar-tensor theories where a scalar field dominates the dynamics of the Universe at late times, in particular we highlight Horndeski and beyond-Horndeski models~\cite{horndeski:horndeskitheory, deffayet:horndeskitheory, kobayashi:horndeskitheory, gleyzes:beyondhorndeskitheory, kobayashi:horndeskireview}. 
Despite having been constrained by early and late time cosmological probes, see, e.g., ref.~\cite{bellini:constraintsonmg}, and by the remarkable binary neutron star event~\cite{creminelli:bnsconstraintsonmg, ezquiaga:bnsconstraintsonmg, baker:bnsconstraintsonmg}, the parameter space in this class of models can still allow for interesting deviations from General Relativity at late times. 
Following the spirit of the Effective Field Theory of Dark Energy~\cite{creminelli:eftofde, gubitosi:eftofde, bloomfield:eftofdeI, gleyzes:eftofde, bloomfield:eftofdeII, piazza:eftofde, bellini:eftofde}, we choose to conveniently parametrize deviations from General Relativity instead of starting from a well defined Lagrangian density. 
In particular, at background level we impose a~$\Lambda$CDM expansion history, while at perturbative level we allow for deviations by introducing two free functions of redshift~$z$ and scale~$k$. 
Between many different parametrization choices~\cite{ade:planckmodifiedgravity}, we choose the~$\mu-\eta$ one given by~\cite{amendola:mgparametrization}
\begin{equation}
    \begin{aligned}
        k^2\Psi(k,z) &= -4\pi G a^2 \mu(k,z) \bar{\rho} D,\\
        \Phi(k,z) &= \eta(k,z) \Psi(k,z), \\
    \end{aligned}
\label{eq:mu_eta_parametrization}
\end{equation}
where the Bardeen potentials~$\Phi,\Psi$ are defined by the line element
\begin{equation}
    ds^2 = a^2\left[-(1+2\Psi)d\tau^2 + (1-2\Phi)\delta^K_{ij} dx^idx^j \right],
\end{equation}
the conformal time and spatial coordinates are~$\tau$ and~$x^j$, respectively, $a$ is the scale factor, $D = \delta + 3\mathcal{H}(1+w)\theta/k^2$ is the gauge-invariant energy density fluctuation, $\delta$ is the energy density fluctuation, $\theta$ is the velocity divergence, $\mathcal{H}=a'/a$ is the Hubble expansion rate in conformal time, $w$ is the equation of state of the Universe, $G$ is Newton constant. 
Regarding the $\mu-\eta$ Modified Gravity functions, the former affects the growth of structure by effectively changing the strength of Newton constant, while the latter changes the relation between the metric potentials.

Regarding the form of~$\mu$ and~$\eta$, several authors have considered a scale-independent form for these functions; however, in total generality, Modified Gravity theories often have additional scales appearing both at linear~\cite{bellini:eftofde} and nonlinear~\cite{khoury:chameleonI, khoury:chameleonII, hinterbichler:symmetron, brax:dilaton, babichev:kmouflage, vainshtein:vainshtein} level. 
We build on the redshift-dependent parametrization introduced in ref.~\cite{zhao:muetaparametrization} and we complement it with an additional scale dependence to account phenomenologically for the presence of an additional scale~$k_\mathrm{mg}$ in the theory as~\cite{alonso:screening, spuriomancini:screening}
\begin{equation}
    \begin{aligned}
        \mu(k,z) &= 1 + \frac{\mu_0-1}{2}\left[1 - \tanh\frac{z-z_\mathrm{th}}{\Delta z_\mathrm{th}}\right]e^{-\frac{1}{2}(k/k_\mathrm{mg})^2}, \\
        \eta(k,z) &= 1 + \frac{\eta_0-1}{2}\left[1 - \tanh\frac{z-z_\mathrm{th}}{\Delta z_\mathrm{th}}\right]e^{-\frac{1}{2}(k/k_\mathrm{mg})^2},
    \end{aligned}
\label{eq:Modified_Gravity_mu_eta}
\end{equation}
where~$\mu_0$, $\eta_0$, $z_\mathrm{th}$, $\Delta z_\mathrm{th}$ and~$k_\mathrm{mg}$ are constant parameters. 
The parameters~$z_\mathrm{th}$ and~$\Delta z_\mathrm{th}$ control when and how fast we transition from the General Relativity to the Modified Gravity regime. 
The General Relativity limit is recovered for~$\mu_0=\eta_0=1$, hence when~$\mu=\eta=1$. 
Moreover, this parametrization allows to recover the~$\Lambda$CDM limit in the Early Universe~($z\gg z_\mathrm{th}$) or at small scales~($k\gg k_\mathrm{mg}$) independently from the value of~$\mu_0$ and~$\eta_0$, while, at the same time, allowing for deviations at late time and large scales. 
Following refs.~\cite{raccanelli:radiosurveysI, raccanelli:radiosurveysII} we choose~$z_\mathrm{th}=6$, $\Delta z_\mathrm{th}=0.05$. 
We choose as benchmark scale $k_\mathrm{mg} = 10^{-3}$, $10^{-2}$ and~$10^{-1}\ \mathrm{Mpc}^{-1}$ to study the dependence of the constraints on the extra scale of the theory. 
The different parametrizations have been implemented in~\texttt{Multi\_CLASS}, and the code has been validated against the results of refs.~\cite{baker:modifiedgravity, sakr:mgclassII} obtained using the~\texttt{MGCLASS II} code (more details are included in appendix~\ref{app:MG_Multi_CLASS}). 

In alternative, other popular Modified Gravity models involve the existence of extra dimensions, as the Dvali-Gabadadze-Porrati (DGP) model~\cite{dvali:dgpmodel}: this 1-parameter extension of~$\Lambda$CDM describes the Universe as a 4D brane embedded in a 5D Minkowski space.
While matter is confined on the brane, gravity propagates in the extra dimension above the scale~$r_c$, which represents the cross-over scale between the 4D and 5D behaviour. 
In this work we focus on only one of the two branches of the theory~\cite{deffayet:dgpmodel}, the so-called ``normal branch'' (hereafter denoted as nDGP), since the ``self-accelerating branch'' has been showed to be unstable~\cite{luty:dgpinstability, charmousis:dgpinstability, koyama:dgpinstability}. 
This simple model has an incredibly rich phenomenology, allowing for potential deviations from GR at the background, linear and non-linear level~\cite{lombriser:dgpconstraints, raccanelli:dgpconstraints, xu:dgpconstraints, barreira:dgpconstraints, piga:dgpconstraints, hernandezaguayo:dgpnonlinearscales}. 
However, in this work we focus on the model proposed in ref.~\cite{schmidt:dgpmodel}, which leaves unaltered the expansion history of the Universe but changes the growth of structure. 
Despite having a substantially different theoretical framework from the case of scalar-tensor theories, also this model can be described at perturbative level using the parametrization given in equation~\eqref{eq:mu_eta_parametrization} by imposing~\cite{lue:mudgpmodel, koyama:mudgpmodel}
\begin{equation}
    \mu(z) = 1 + \frac{1}{3\beta}, \qquad \eta(z) = \frac{1 - \frac{1}{3\beta}}{1 + \frac{1}{3\beta}},
\label{eq:nDGP_mu_eta}
\end{equation}
where
\begin{equation}
    \beta(z) = 1 + \frac{H}{H_0\sqrt{\Omega_\mathrm{rc}}} \left[1 + \frac{1}{3}\frac{d\log H}{d\log a} \right] = 1 + \frac{H}{H_0\sqrt{\Omega_\mathrm{rc}}} \left[1 + \frac{\dot{H}}{3H^2} \right], \qquad \Omega_\mathrm{rc} = \frac{1}{4r_c^2H_0^2},
\label{eq:nDGP_beta}
\end{equation}
and~$H=\dot{a}/a$ and~``$\ \dot{\ }\ $'' represents a derivative with respect to cosmic time. 
In this case we do not include any additional scale dependence in the Modified Gravity functions. 

In conclusion, in the case where we investigate dark matter properties, we consider as cosmological parameters for the forecast
\begin{equation}
    \{\theta_\alpha\} = \{ \omega_\mathrm{cdm}, h, n_s, b^\mathrm{eff}_\mathrm{g}, b^\mathrm{eff}_\mathrm{GW} \},
\label{eq:LambdaCDM_parameters}
\end{equation}
where~$\omega_\mathrm{cdm}$ is the cold dark matter physical density, $h$ is the Hubble parameter, $n_s$ is the spectral index of scalar perturbations and~$b^\mathrm{eff}_\mathrm{g}, b^\mathrm{eff}_\mathrm{GW}$ are the effective bias parameters for galaxies and GWs. 
The fiducial value of GW effective bias depends on binary formation channels, as explained in section~\ref{sec:tracers}. 
On the other hand, in the case where we forecast sensitivity for dark energy models, we consider 
\begin{equation}
    \{\theta_\alpha\} = \{ \omega_\mathrm{cdm}, h, n_s, \mu_0, \eta_0, b^\mathrm{eff}_\mathrm{g}, b^\mathrm{eff}_\mathrm{GW} \},
\end{equation}
and 
\begin{equation}
    \{\theta_\alpha\} = \{ \omega_\mathrm{cdm}, h, n_s, \Omega_\mathrm{rc}, b^\mathrm{eff}_\mathrm{g}, b^\mathrm{eff}_\mathrm{GW} \}.
\end{equation}
for the Modified Gravity and nDGP cases, respectively.
In the latter two cases, the GW effective bias refers to the case of astrophysical BBHs.\footnote{
In principle we could fully fixed the GW bias when investigating the phenomenology of dark energy models instead of introducing a GW effective bias in the Fisher matrix analysis.
We choose to maintain the GW effective bias as a parameter of the analysis since we consider this choice a reasonable trade-off between assuming perfect knowledge of the GW bias, which results in tighter constraints of the Modified Gravity parameters, and a real data analysis, where the most conservative approach would very likely require treating the bias in each individual bin as a separate parameter to be constrained by observations, resulting in looser constraints.
Moreover, we also want to check at the end of the astrophysical-versus-primordial-BHs part of the analysis whether the lack of perfect knowledge of Gravity at large scales significantly affects our ability to test for BH origin.} 
The fiducial value of the standard cosmological parameters are~$\{ \omega_\mathrm{cdm}, h, n_s \} = \{0.12038, 0.67556, 0.9619 \}$, while for the Modified Gravity models we use~$\{\mu_0,\eta_0\}=\{0.87,1.3\}$ and for nDGP we choose~$\{ \Omega_\mathrm{rc} \} = \{ 0.2 \}$.


\section{Tracers}
\label{sec:tracers}

In this section we present the specifics of the two kinds of galaxy survey (SW or DL, \S~\ref{subsec:galaxy_surveys}) and the details on the different astrophysical (\S~\ref{subsec:astro_bbh}) or primordial (\S~\ref{subsec:late_time_pbbh}, \ref{subsec:early_time_pbbh}) BH binaries. 
Regarding PBHs, we consider binaries that form either at late or early times, since they trace LSS in well defined ways. 
Including other formation channels in specific environment, see, e.g., ref.~\cite{valbusadallarmi:pbhinclusters}, or formation channels that include also neutron stars~\cite{sasaki:pbhneutronstarmergers} is left for future work. 
Finally, in \S~\ref{subsec:full_bbh_population} we show how to combine GW events coming from different binary models in a single framework. 

Binary models have been studied using the external modules of~\texttt{CLASS\_GWB}~\cite{bellomo:classgwb}, and the detectability of these sources has been done both for second and third generation GW detector networks, as detailed in appendix~\ref{app:detector_networks}. 
In particular, we consider as detector networks:
\begin{itemize}
    \item HLVIK, a second generation network made by Hanford and Livingston LIGO, Virgo, KAGRA and LIGO India L-shaped interferometers;
    \item ET2CE, a third generation network made by Einstein Telescope (triangle-shaped interferometer) and two Cosmic Explorer (L-shaped interferometers).
\end{itemize}
We assume all the detector to work at their design sensitivity, and individual events are considered ``detected'' if their Signal-to-Noise ratio for the network is~$\rho \geq 12$, see, e.g.,~ref.~\cite{bellomo:classgwb} and references therein. 
Errors on the redshift determination are of the order of~$\delta z/z \simeq \delta d_L/d_L \simeq 3/\rho \lesssim 0.3$~\cite{calore:gwxlss, delpozzo:reshifterror} and are discussed further in appendix~\ref{app:detector_networks} along with typical average and maximum resolution of the two detector networks. 
For any effective purpose, both second and third generation detector networks cover the entire sky (even if with different sensitivities at different times), therefore for GWs we assume~$f_\mathrm{sky}=1.0$. 
In the next subsections we provide fitting formulas to allow interested readers to reproduce our results, without claiming that they provide any particular insight on the properties of the tracers.\footnote{
It should be noted that the reported fitting formula are valid only in the following redshift ranges: $[0,2]$ and~$[0,9]$ for second and third generation astrophysical BH binaries, $[0,2]$ and~$[0,7]$ for second and third generation late time PBH binaries, and $[0,2]$ and~$[0,7]$ for second and third generation early time PBH binaries, respectively.} 


\subsection{Galaxy surveys}
\label{subsec:galaxy_surveys}

First we consider a DL survey that observes up to~$326\times 10^6$ galaxies over~$8000\ \mathrm{deg}^2$ ($f_\mathrm{sky}=0.2$), covering the~$[0.5,4.1]$ redshift range. 
For the sake of this analysis we assume data is grouped into~$18$ redshift bins of half-width~$\Delta z=0.1$. 
On the other hand, for the SW survey we consider a mission that covers the entire sky ($f_\mathrm{sky}=0.7$) in the redshift range~$[0.2,4.2]$. 
In this case we bin galaxies into~$10$ redshift bins with half-width~$\Delta z=0.2$.
These specifics are inspired by those of SPHEREx~\cite{dore:spherexwhitepaperI, dore:spherexwhitepaperII, dore:spherexwhitepaperIII}.\footnote{
\url{https://github.com/SPHEREx/Public-products}.
}
In both cases we consider tophat window functions when binning the galaxy catalogs.

The redshift distributions of the galaxy populations for both surveys can be parametrized as
\begin{equation}
    \frac{d^2N_\mathrm{g}}{dzd\Omega} = \mathcal{A}\left(\frac{z}{z_0}\right)^\alpha e^{-(z/z_0)^\beta},
\label{eq:dNdzdOmega_galaxies}
\end{equation}
where for the DL survey~$\mathcal{A}=194575\ \mathrm{gal/deg^2}$, $z_0 = 0.07$, $\alpha = 0.34$ and~$\beta = 0.42$; while for SW one we consider~$\mathcal{A}=25509\ \mathrm{gal/deg^2}$, $z_0 = 0.09$, $\alpha = 1.75$ and~$\beta = 0.69$, which take into account the contribution of each sample.

The bias for the DL survey is approximated as~$b_\mathrm{g} = 0.84/D(z)$, where~$D(z)$ is the linear growth factor normalized to~$1$ at redshift~$z=0$, while for SW one we consider a parametric form
\begin{equation}
    b_\mathrm{g}(z)= b_0 + b_1 z + b_2 z^2,
\end{equation}
where~$b_0=0.53$, $b_1=1.59$ and~$b_2=-0.08$.
Therefore, the effective galaxy bias for these two surveys reads as~$b^\mathrm{eff}_\mathrm{g}=1.76$ and~$b^\mathrm{eff}_\mathrm{g}=1.69$ for DL and SW, respectively. 
Regarding the magnification bias~$s_\mathrm{g}$, since this is a preliminary study and no accurate data is available, we choose~$s_\mathrm{g} = 0.6$ for both surveys. 
Finally, the evolution bias~$f^\mathrm{evo}_\mathrm{g}$ is inferred from the galaxy redshift distribution given in equation~\eqref{eq:dNdzdOmega_galaxies}.


\subsection{Astrophysical black hole binaries}
\label{subsec:astro_bbh}

First of all we consider GWs emitted by coalescing BHs generated as a result of stellar evolution. 
These BHs are expected to live in galaxies, where the star formation history is more intense, hence they are expected to trace LSS in a similar (yet not identical) way to what galaxies do. 
We create catalogs\footnote{Typically multiple catalogs of~$10^5$ events are created to test consistency of the inferred properties such as the GW redshift distribution. Given our fiducial model, each catalog covers an observation time of~$T_\mathrm{obs}=2.92$ years.} of GW events following the model and the procedure presented in ref.~\cite{bellomo:classgwb}, hence, for the sake of conciseness, we will not repeat all the details in this article. 
Suffice to say that every GW event is described by its time of arrival at Earth, merger redshift, binary formation redshift, dark matter halo mass both at formation and at merger, masses and spins of the compact objects, sky localization of the event, inclination angle of the binary and polarization of the signal. 
Assuming that all the BH binaries are astrophysical in origin, the LIGO-Virgo-KAGRA Collaboration measured a local merger rate of~$\bar{R}^\mathrm{LVK}_0\simeq 20\ \mathrm{Gpc^{-3}yr^{-1}}$~\cite{abbott:gwtc3properties}.

Given the catalogs, we compute the observed GW redshift distribution both for second and third generation detector networks and we normalized it to~$T_\mathrm{obs}=10$ years of observation time. 
We parametrize the GW redshift distribution as in equation~\eqref{eq:dNdzdOmega_galaxies}, finding~$\{ \mathcal{A}^\mathrm{2G}, z^\mathrm{2G}_0, \alpha^\mathrm{2G}, \beta^\mathrm{2G} \} = \{84.26\ \mathrm{GW/sr}, 0.04,  2.79, 0.68\}$ and~$\{\mathcal{A}^\mathrm{3G}, z^\mathrm{3G}_0, \alpha^\mathrm{3G}, \beta^\mathrm{3G} \} = \{1.35 \times 10^{-7}\ \mathrm{GW/sr}, 2.02 \times 10^{-3}, 6.12, 0.41\}$ for second and third generation detectors, respectively.  
The bias of each individual GW event is computed starting from the dark matter halo mass at merger and by using the halo bias model, as proposed in refs.~\cite{bertacca:gwbprojectioneffects, bellomo:classgwb}. 
The biases are then averaged in every redshift bin and the overall bias function is parametrized as
\begin{equation}
    b_\mathrm{GW}(z)= b_0 + b_1 z+ b_2 z^2+ b_3 z^3,
\label{eq:GW_bias}
\end{equation}
where~$\{ b^\mathrm{2G}_0, b^\mathrm{2G}_1, b^\mathrm{2G}_2, b^\mathrm{2G}_3 \} = \{0.86,0.12,0.75,-0.27\}$ and~$\{ b^\mathrm{3G}_0, b^\mathrm{3G}_1, b^\mathrm{3G}_2, b^\mathrm{3G}_3 \} = \{0.65,1.45,-0.14,0.01\}$ for second and third generation detector networks, respectively. 
The effective GW bias is given by~$b^\mathrm{2G,eff}_\mathrm{GW}=1.29$ and~$b^\mathrm{3G,eff}_\mathrm{GW}=2.86$, in the redshift range probed by the DL survey, and by~$b^\mathrm{2G,eff}_\mathrm{GW}=1.14$ and~$b^\mathrm{3G,eff}_\mathrm{GW}=2.84$ in the SW one.
Regarding the GW magnification bias
\begin{equation}
    s_\mathrm{GW}(z) = - \left. \frac{d\log_{10}\frac{d2N_\mathrm{GW}}{dzd\Omega}}{d\rho} \right|_{\rho=12}
\end{equation}
we confirm the result first presented in ref.~\cite{scelfo:gwxlssI} that~$s^\mathrm{2G}_\mathrm{GW}\simeq [0.0,0.6]$ and~$s^\mathrm{3G}_\mathrm{GW}\simeq 0$ for second and third generation of detectors, respectively (see also figures~\ref{fig:BH_2G_mixed_models} and~\ref{fig:BH_mixed_models}). 
The GW evolution bias, as for galaxies, is computed from the GW redshift distribution.


\subsection{Late time primordial black hole binaries}
\label{subsec:late_time_pbbh}

PBHs inside virialized dark matter halos can form binaries at late times through a direct capture process of unbound compact objects~\cite{bird:pbhlatetimebinaries, clesse:pbhlatetimebinaries}, similarly to standard astrophysical BHs. 
In this article we follow the approach of ref.~\cite{bird:pbhlatetimebinaries} to compute the intrinsic merger rate density for PBH late time binaries as
\begin{equation}
    \bar{R}^\mathrm{LPBH}_\mathrm{m}(z) = \int_{t_{d,\mathrm{min}}}^{t_{d,\mathrm{max}}} dt_d p(t_d) \bar{R}_\mathrm{bf}(z_f),
\label{eq:late_time_merger_rate_density}
\end{equation}
where~$t_d$ is the time delay between merger and binary formation, $p(t_d)$ is the time delay probability density function, $z_f$ is the binary formation redshift and $\bar{R}_\mathrm{bf}$ is the intrinsic binary formation rate density (see appendix~\ref{app:lpbh_binary_formation} for details on its computation). 
By simulating catalogs of binaries in a similar fashion to what we do to create GW events, we derived for the first time the expected time delay probability distribution function and, as we show in appendix~\ref{app:lpbh_binary_formation}, we can approximate it as~$p(t_d) \propto t^{-1}_d$. 
However, since~$t_{d,\mathrm{max}} \simeq t(z)$, where~$t(z)$ is the age of the Universe at redshift~$z$, the time delay is non-negligible.
The local merger rate of~$\bar{R}^\mathrm{LPBH}_\mathrm{m,0}\simeq 2\ \mathrm{Gpc^{-3}yr^{-1}}$ to be compared to the (overestimated) local merger rate~$\bar{R}^\mathrm{LPBH}_\mathrm{m,0}\simeq 9.5\ \mathrm{Gpc^{-3}yr^{-1}}$ obtained without including the time delay effect.

As for astrophysical BHs in the previous section, we use the catalog approach to derive the late time PBH merging binary redshift distribution and bias.
Parametrizing again the GW redshift distribution with equation~\eqref{eq:dNdzdOmega_galaxies}, we find~$\{ \mathcal{A}^\mathrm{2G}, z^\mathrm{2G}_0, \alpha^\mathrm{2G}, \beta^\mathrm{2G} \} = \{119.29\ \mathrm{GW/sr}, 0.20,  2.40, 1.05\}$ and~$\{\mathcal{A}^\mathrm{3G}, z^\mathrm{3G}_0, \alpha^\mathrm{3G}, \beta^\mathrm{3G} \} = \{0.2 \ \mathrm{GW/sr}, 5 \times 10^{-3}, 2.42, 0.33\}$ for second and third generation detectors, respectively.
For the bias we follow the same methodology used for astrophysical BHs and, parametrizing the bias as in equation~\eqref{eq:GW_bias}, we find~$\{ b^\mathrm{2G}_0, b^\mathrm{2G}_1, b^\mathrm{2G}_2, b^\mathrm{2G}_3 \} = \{0.60,-0.01,3 \times 10^{-3},1 \times 10^{-3}\}$ and~$\{ b^\mathrm{3G}_0, b^\mathrm{3G}_1, b^\mathrm{3G}_2, b^\mathrm{3G}_3 \} = \{0.60,-0.01,6 \times 10^{-3},-2 \times 10^{-4}\}$ for second and third generation detector networks, respectively.
The effective GW bias is given by~$b^\mathrm{2G,eff}_\mathrm{GW}=0.59$ and~$b^\mathrm{3G,eff}_\mathrm{GW}=0.61$ in both the redshift ranges probed by the DL and SW surveys.
Also in this case, we find that~$s^\mathrm{2G}_\mathrm{GW}\simeq [0.0,0.6]$ and~$s^\mathrm{3G}_\mathrm{GW}\simeq 0$ and the GW evolution bias is calculated from the GW redshift distribution.


\subsection{Early time primordial black hole binaries}
\label{subsec:early_time_pbbh}

Alternatively, PBH binaries can form before matter-radiation equality~\cite{nakamura:earlytimebinaries, ioka:earlytimebinaries, sasaki:earlytimebinaries}. 
Thanks to tidal forces generated by other PBHs, the binary components avoid an head-to-head collision and start their inspiralling motion, eventually merging at later times.
Numerous authors analysed both the binary creation stage and the binary evolution history, finding that overall the merger rate of such PBH binaries can be modelled as~\cite{alihaimoud:earlytimebinaries, raidal:earlytimebinariesI, raidal:earlytimebinariesII}
\begin{equation}
    \bar{R}^\mathrm{EPBH}_\mathrm{m}(z) = f^{53/37}_\mathrm{PBH} \left[\frac{t_0}{t(z)}\right]^{\frac{34}{37}} \int dM_1 dM_2\ \mathcal{A}_\mathrm{m} \left(\frac{M}{60\ M_\odot}\right)^{\frac{2}{37}} \left(\frac{\mu}{15\ M_\odot}\right)^{-\frac{34}{37}} \frac{d\Phi_1}{dM_1} \frac{d\Phi_2}{dM_2},
\end{equation}
where~$t_0=t(0)$ is the age of the Universe, and~$\mathcal{A}_\mathrm{m}$ is an amplitude factor that accounts for effects connected to the distribution of initial binaries~\cite{raidal:earlytimebinariesII, ballesteros:earlybinaries, deluca:earlybinaries}, cosmological perturbations~\cite{garriga:earlytimebinaries}, interaction with surrounding environment~\cite{hayasaki:earlytimebinaries}, and many more. 
Typical values of the amplitude factor are of order~$\mathcal{O}(10^5)\ \mathrm{Gpc^{-3}yr^{-1}}$, which, if taken at face value, would constraint PBH abundance to be of order~$f_\mathrm{PBH}\sim\mathcal{O}(10^{-3})$~\cite{deluca:pbhabundance, hall:pbhabundance, hutsi:pbhabundance}. 
However, numerical simulations seem to suggest that binary disruption due to three body encounters might be more frequent than what was previously thought~\cite{jedamzik:earlytimebinariesI, jedamzik:earlytimebinariesII}, and that the real value of the amplitude factor is of order~$\mathcal{A}_\mathrm{m} \sim \mathcal{O}(10)\ \mathrm{Gpc^{-3}yr^{-1}}$ in the tens of solar masses range, making PBH a viable candidate for dark matter.
In the following we fix as benchmark value~$\mathcal{A}_\mathrm{m}=18\ \mathrm{Gpc^{-3}yr^{-1}}$: in this fashion, when~$f_\mathrm{PBH}=1$, we have that late and early time PBH binaries completely saturate the measured local merger rate~$\bar{R}^\mathrm{LVK}_0$.

Also for early time PBH binaries we create catalogs of events and assess how many are detected by different GW detector networks.
Parametrizing the GW redshift distribution according to equation~\eqref{eq:dNdzdOmega_galaxies}, we find~$\{ \mathcal{A}^\mathrm{2G}, z^\mathrm{2G}_0, \alpha^\mathrm{2G}, \beta^\mathrm{2G} \} = \{133.75\ \mathrm{GW/sr}, 0.09,  3.41, 0.84\}$ and~$\{\mathcal{A}^\mathrm{3G}, z^\mathrm{3G}_0, \alpha^\mathrm{3G}, \beta^\mathrm{3G} \} = \{9.40\ \mathrm{GW/sr}, 0.03, 2.85, 0.41\}$ for second and third generation detectors, respectively. 
Since early time PBH binaries formed during the radiation dominated era and they follow the subsequent dark matter evolution, we assume they trace very well the underlying matter distribution, hence we assume their bias to be equal to~$b_\mathrm{GW}=1$, and therefore also the effective bias is~$b^\mathrm{eff}_\mathrm{GW}=1$. 
As for astrophysical BH and late time PBH binaries, we find~$s^\mathrm{2G}_\mathrm{GW}\simeq [0.0,0.6]$ and~$s^\mathrm{3G}_\mathrm{GW}\simeq 0$ for second and third generation detectors and we calculate the GW evolution bias from the GW redshift distribution.


\subsection{Complete black hole binary population}
\label{subsec:full_bbh_population}

So far we consider cases where only a single BH binary population contributes to observations, however in reality all these populations contribute to the totality of detected events. 
Hence the true functions characterizing the two-point statistics of GW are the total number density
\begin{equation}
    \frac{d^2N^\mathrm{tot}_\mathrm{GW}}{dzd\Omega} = \frac{d^2N_\mathrm{ABH}}{dzd\Omega} + \frac{d^2N_\mathrm{LPBH}}{dzd\Omega} + \frac{d^2N_\mathrm{EPBH}}{dzd\Omega},
\end{equation}
and, by definition, total bias, magnification bias and evolution bias
\begin{equation}
    \begin{aligned}
        \frac{d^2N^\mathrm{tot}_\mathrm{GW}}{dzd\Omega} b^\mathrm{tot}_\mathrm{GW} &= \frac{d^2N_\mathrm{ABH}}{dzd\Omega} b_\mathrm{ABH} + \frac{d^2N_\mathrm{LPBH}}{dzd\Omega} b_\mathrm{LPBH} + \frac{d^2N_\mathrm{EPBH}}{dzd\Omega} b_\mathrm{EPBH}, \\
        \frac{d^2N^\mathrm{tot}_\mathrm{GW}}{dzd\Omega} s^\mathrm{tot}_\mathrm{GW} &= \frac{d^2N_\mathrm{ABH}}{dzd\Omega} s_\mathrm{ABH} + \frac{d^2N_\mathrm{LPBH}}{dzd\Omega} s_\mathrm{LPBH} + \frac{d^2N_\mathrm{EPBH}}{dzd\Omega} s_\mathrm{EPBH}, \\
        \frac{d^2N^\mathrm{tot}_\mathrm{GW}}{dzd\Omega} f^\mathrm{evo,tot}_\mathrm{GW} &= \frac{d^2N_\mathrm{ABH}}{dzd\Omega} f^\mathrm{evo}_\mathrm{ABH} + \frac{d^2N_\mathrm{LPBH}}{dzd\Omega} f^\mathrm{evo}_\mathrm{LPBH} + \frac{d^2N_\mathrm{EPBH}}{dzd\Omega} f^\mathrm{evo}_\mathrm{EPBH}.
    \end{aligned}  
\label{eq:mixed_model_bias_functions}
\end{equation}

The relative importance of PBH with respect to astrophysical BHs depends on the~$f_\mathrm{PBH}$ and~$\mathcal{A}_\mathrm{m}$ parameters. 
We investigate the PBH parameter space given by~$\mathcal{A}_\mathrm{m}\in [0,18]$ and~$f_\mathrm{PBH}\in [0,1]$, where, at every point, the local merger rate of astrophysical BHs is defined by
\begin{equation}
    \bar{R}^\mathrm{ABH}_0 = \bar{R}^\mathrm{LVK}_0 - \bar{R}^\mathrm{LPBH}_0 - \bar{R}^\mathrm{EPBH}_0 = \bar{R}^\mathrm{LVK}_0 - 2f_\mathrm{PBH}^2\ - \mathcal{A}_\mathrm{m} f^{53/37}_\mathrm{PBH} \ \mathrm{Gpc^{-3}yr^{-1}},
\label{eq:R0_ABH}
\end{equation}
in order to saturate by definition the measured local merger rate.

\begin{figure}[t]
    \centerline{
    \includegraphics[width=\columnwidth]{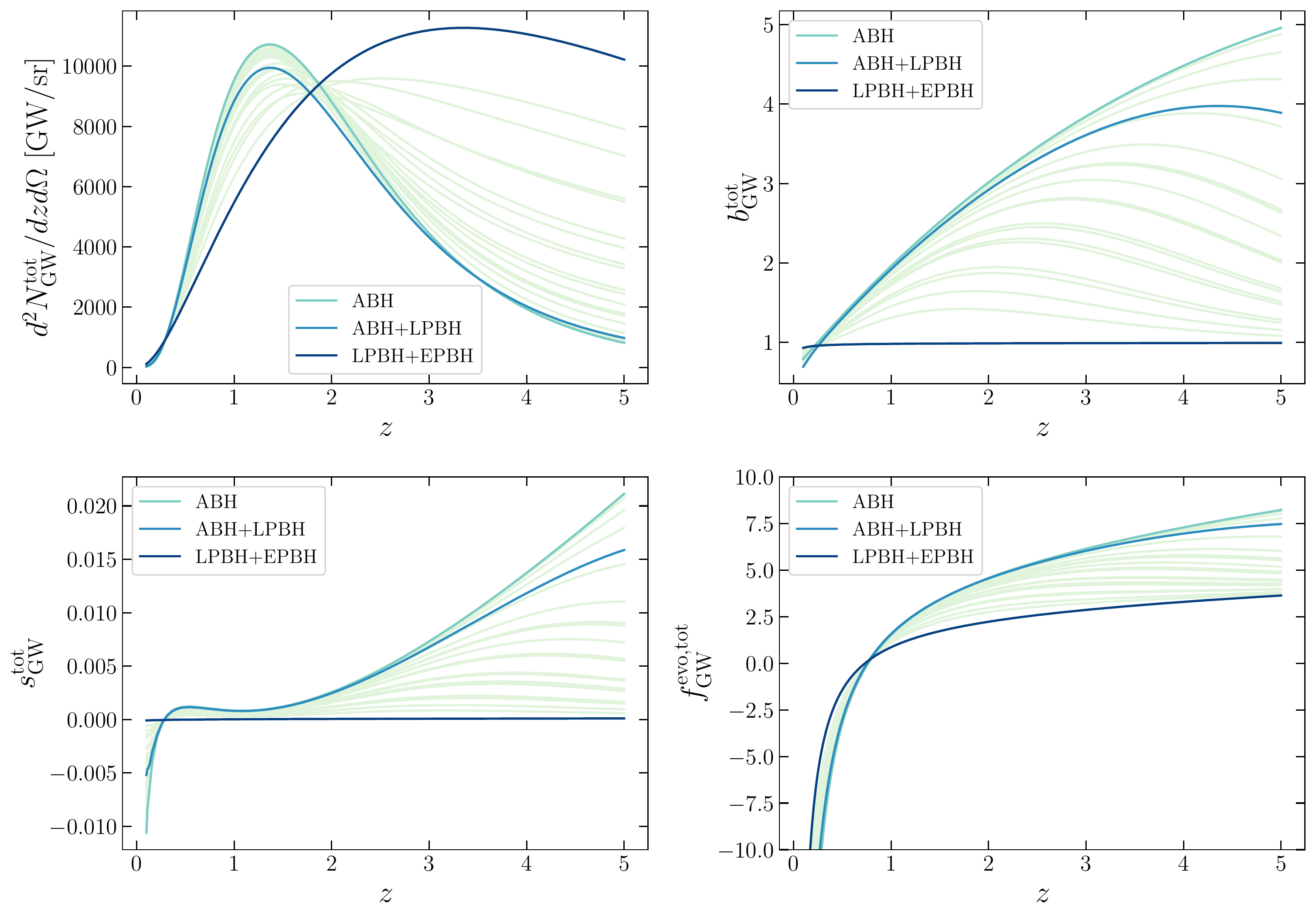}}
\caption{GW mixed population redshift distribution (\textit{top left panel}), bias (\textit{top right panel}), magnification bias (\textit{bottom left panel}) and evolution bias (\textit{bottom right panel}) for different values of~$f_\mathrm{PBH}$ and~$\mathcal{A}_m$.
In this plot we consider only events detected by a third generation GW detector network.
The \textit{light blue}, \textit{blue} and \textit{dark blue} lines correspond to the astrophysical BHs only ($f_\mathrm{PBH}=\mathcal{A}_\mathrm{m}=0$), mixed astrophysical and late time PBH ($f_\mathrm{PBH}=1,\ \mathcal{A}_\mathrm{m}=0$), and PBHs only ($f_\mathrm{PBH}=1,\ \mathcal{A}_\mathrm{m}=18$) scenarios, respectively.
In the mixed astrophysical and late time PBH scenario, late time PBHs contribute to approximately~$6\%$ of the total number of observed events.
The \textit{green} lines refer to mixed models where different populations contribute to the total number of detected events.} 
\label{fig:BH_mixed_models}
\end{figure}

We show in figure~\ref{fig:BH_mixed_models} the total redshift distribution, bias, magnification bias and evolution bias for different models spanning the~$f_\mathrm{PBH}-\mathcal{A}_\mathrm{m}$ parameter space for a third generation detector network, along with three reference scenarios corresponding to the astrophysical BHs only ($f_\mathrm{PBH}=\mathcal{A}_\mathrm{m}=0$), mixed astrophysical and late time PBH ($f_\mathrm{PBH}=1,\ \mathcal{A}_\mathrm{m}=0$), and PBHs only ($f_\mathrm{PBH}=1,\ \mathcal{A}_\mathrm{m}=18$) cases.
In the astrophysical BH only scenario we observe a redshift distribution peaked~$z\simeq 1.5$, when also the cosmic star formation rate peaks.
Increasing the abundance of PBHs slightly alters the distribution, if only late time PBH binaries are included, or significantly shifts it at higher redshift of order~$z\simeq 3.5$, when early time PBH binaries are included.
At the same time, the inclusion of PBH significantly lowers the total bias of the population, especially at high redshift, while affecting only partially the total magnification and evolution bias.
The analogue figure for GW events detected by a second generation detector network is reported in appendix~\ref{app:mixed_BBH_2G}.

Finally, we note that, even keeping fixed the total local merger rate to the measured value, the total number of detected GW events changes due to the different merger rate redshift dependence of different models.
Since the average number of detected events is~$N_\mathrm{ABH}\simeq 33000\ \mathrm{GW\, yr^{-1}}$ when $f_\mathrm{PBH}=\mathcal{A}_\mathrm{m}=0$ (astrophysical BHs only scenario) or~$N_\mathrm{PBH}\simeq 43000\ \mathrm{GW\, yr^{-1}}$ when $f_\mathrm{PBH}=1,\ \mathcal{A}_\mathrm{m}=18$ (PBH only scenario, $41000$ early time and~$2000$ late time events), over the span of~$T_\mathrm{obs}=10\ \mathrm{yr}$ we would observe approximately between~$2.9\times 10^5$ and~$4.3\times 10^5$ GW events, depending on how much each population contributes to the total.
These estimates are considerably influenced by the shape of the merger rate at high redshift, which is currently unknown, however we can already use them to infer how much the shot noise changes in our forecast.
In particular we can see that it would change at most by a factor~$1.5$ over the entire~$f_\mathrm{PBH}-\mathcal{A}_\mathrm{m}$ parameter space.
The astrophysical BH model we adopt can be seen as conservative compared to existing estimates that predict up to~$N_\mathrm{ABH}\simeq 7-8\times 10^4$ observed events per year~\cite{iacovelli:gw3genforecast}, hence also our errors and detectability prospects can be interpreted as conservative.


\section{Testing extended cosmologies}
\label{sec:results}

In this section we present the capability of GWs to test cosmological model that deviates from the standard~$\Lambda$CDM either in the dark matter or in the dark energy sectors. 
We present results for GW alone and in synergy with galaxy surveys, in particular we discuss whether a DL or a SW survey are more effective in constraining our set of extended cosmologies.
In order to draw fair conclusions regarding both kind of surveys, first we report in table~\ref{tab:cosmo_constraints} errors on standard cosmological parameters: our results suggest that the DL survey is equal to or even a factor two more efficient than the SW survey in constraining them.

\begin{table}[ht]
    \centerline{
    \begin{tabular}{|c|c|c|c|c|}
    \hline
    Survey & $\sigma_{n_s}$ & $\sigma_{h}$ & $\sigma_{\omega_\mathrm{cdm}}$  & $\sigma_{b_\mathrm{g}}$ \\
    \hline
    \hline
    SW & $0.017$ & $0.019$ & $0.005$ & $0.04$  \\
    DL & $0.013$ & $0.009$ & $0.003$ & $0.03$  \\
    \hline
    \end{tabular}}
\caption{Constraints on standard cosmological parameters from the DL and SW galaxy surveys. 
Errors are calculated for the Modified Gravity model with $k_\mathrm{mg} = 10^{-3}\ \mathrm{Mpc}^{-1}$, however they are consistent also with the errors found for other choices of~$k_\mathrm{mg}$ and for the nDGP model.}
\label{tab:cosmo_constraints}
\end{table}


\subsection{Constraints on modified gravity theories}
\label{subsec:mg_constraints}

\begin{figure}[t]
    \centerline{
    \includegraphics[width=\columnwidth]{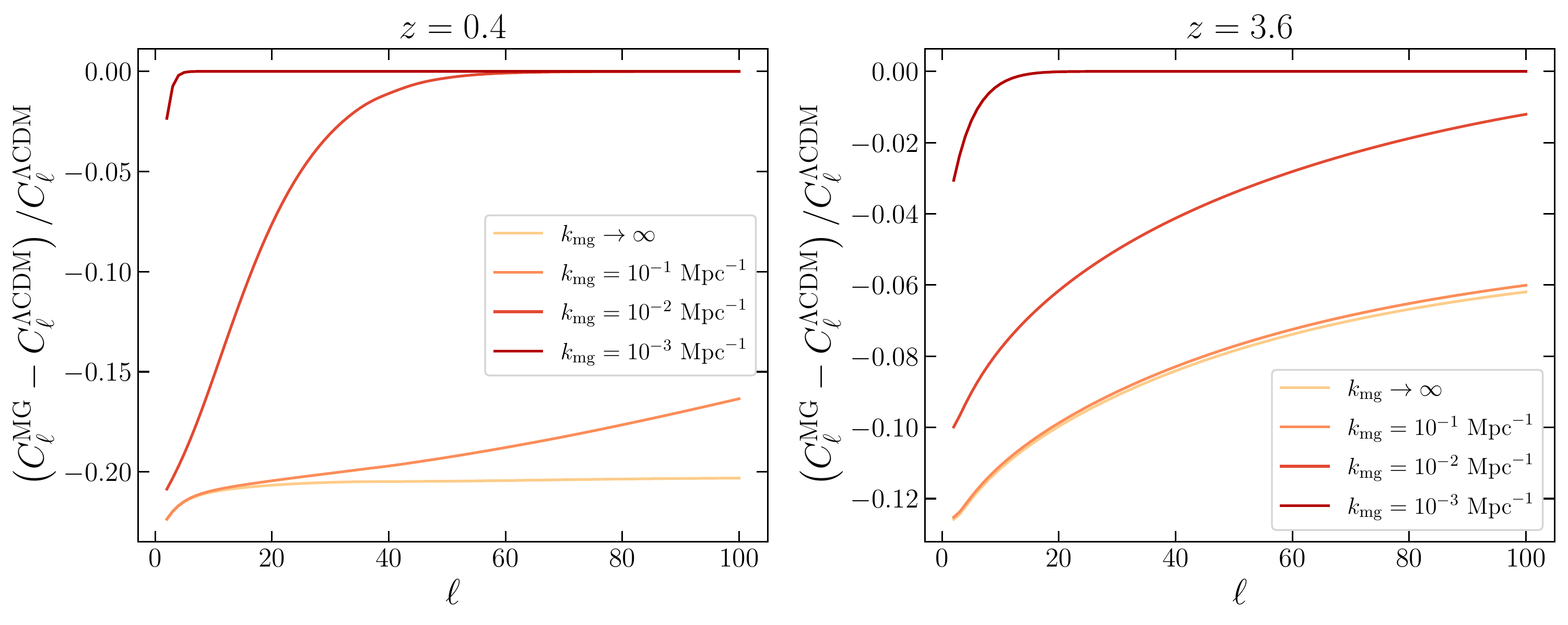}}
\caption{Galaxy angular power spectrum relative difference between Modified Gravity (MG) scenarios with respect to the $\Lambda$CDM case for redshift bins centered at~$z=0.4$ (\textit{left panel}) and~$z=3.6$ (\textit{right panel}). 
The Modified Gravity extra scale dependence is set at~$k_\mathrm{mg}\to\infty$ (\textit{yellow} line) and~$k_\mathrm{mg}=10^{-1}$, $10^{-2}$ and~$10^{-3}\ \mathrm{Mpc}^{-1}$ (\textit{orange}, \textit{red} and \textit{dark red} lines, respectively). 
We show the effect only for galaxies, however also the GW angular power spectrum and the galaxy-GW angular power spectra share a similar phenomenology.}
\label{fig:Clgg_relativedifference}
\end{figure}

First of all, we consider the effect of the additional Modified Gravity scale dependence in the~$\mu(k,z)$ and~$\eta(k,z)$ functions parametrized by~$k_\mathrm{mg}$. 
We present in figure~\ref{fig:Clgg_relativedifference} the relative difference between the angular power spectrum in two different redshift bins in Modified Gravity models with~$k_\mathrm{mg} = 10^{-1}$, $10^{-2}$ and~$10^{-3}\ \mathrm{Mpc}^{-1}$ with respect to the~$\Lambda$CDM case. 
We also show in the same figure the case in which no extra scale dependence is included as a useful comparison obtained: this scenario is obtained by taking the limit~$k_\mathrm{mg}\to\infty$, in which~$\mu$ and~$\eta$ become functions only of redshift. 
The suppression effect showed in figure~\ref{fig:Clgg_relativedifference} can be understood by comparing the Modified Gravity scale~$k_\mathrm{mg}$ to the typical scale associated to each multipole $k\simeq (\ell+1/2)/r(z)$~\cite{kaiser:limberapproximation, loverde:extendedlimberapproximation}, where~$r(z)$ is the comoving distance to the redshift bin centered at~$z$. 
In the cases showed in the figure we have~$r(0.4)\simeq 1.6\ \mathrm{Gpc}$ and~$r(3.6)\simeq 7\ \mathrm{Gpc}$, therefore at multipoles such that~$\ell \gg \ell_\mathrm{mg}(z) = k_\mathrm{mg}r(z)-1/2$ we expect deviations from $\Lambda$CDM to be exponentially suppressed and to recover the~$\Lambda$CDM limit.
In this example the critical multipole at which the two scales are approximately equal is~$\ell_\mathrm{mg}(0.4)\simeq 160,\ 16,\ 1$ ($\ell_\mathrm{mg}(3.6)\simeq 700,\ 70,\ 7$) for~$k_\mathrm{mg} = 10^{-1}$, $10^{-2}$ and~$10^{-3}\ \mathrm{Mpc}^{-1}$, respectively, as it could have been estimated by eye in the figures.

\begin{table}[t]
    \centering
    \begin{tabular}{|c|c|c|c|c|}
    \hline
    Tracers & $k_\mathrm{mg}\ [\mathrm{Mpc}^{-1}]$ & $\sigma_{\mu_0}$ & $\sigma_{\eta_0}$ & $\sigma_{b^\mathrm{eff}_\mathrm{GW}}$ \\
    \hline
    \hline
    \multirow{3}{*}{ET2CE} & $10^{-1}$ & $0.75\ (0.58)$ & $5.40\ (2.99)$ & $2.97\ (1.08)$ \\
                           & $10^{-2}$ & $1.93\ (1.50)$ & $5.86\ (4.14)$ & $2.65\ (0.25)$ \\
                           & $10^{-3}$ & $55.78\ (55.48)$ & $72.53\ (72.00)$ & $1.94\ (0.19)$ \\
    \hline
    \multirow{3}{*}{SW$\times$ET2CE} & $10^{-1}$ & $0.03\ (0.02)$ & $0.15\ (0.06)$ & $0.11\ (0.06)$ \\
                                     & $10^{-2}$ & $0.11\ (0.10)$ & $0.29\ (0.25)$ & $0.09\ (0.06)$ \\
                                     & $10^{-3}$ & $2.29\ (2.29)$ & $2.98\ (2.98)$ & $0.09\ (0.05)$ \\
    \hline
    \multirow{3}{*}{DL$\times$ET2CE} & $10^{-1}$ & $0.03\ (0.02)$ & $0.10\ (0.06)$ & $0.09\ (0.07)$ \\
                                     & $10^{-2}$ & $0.11\ (0.11)$ & $0.29\ (0.24)$ & $0.08\ (0.07)$ \\
                                     & $10^{-3}$ & $3.59\ (3.59)$ & $2.98\ (2.97)$ & $0.08\ (0.07)$ \\
    \hline
    \end{tabular}\\
    \vspace{0.3cm}
    \begin{tabular}{|c|c|c|}
    \hline
    Tracers & $\sigma_{\Omega_\mathrm{rc}}$ & $\sigma_{b^\mathrm{eff}_\mathrm{GW}}$ \\
    \hline
    \hline
    ET2CE           & $4.18\ (2.98)$ & $1.87\ (0.67)$ \\
    SW$\times$ET2CE & $0.11\ (0.02)$ & $0.09\ (0.05)$ \\
    DL$\times$ET2CE & $0.10\ (0.03)$ & $0.08\ (0.06)$ \\
    \hline
    \end{tabular}
\caption{Marginalized errors on Modified Gravity (\textit{upper table}) and nDGP (\textit{lower table}) parameters for single (GW only) and multi-tracer analysis. 
SW and DL refer to ``shallow and wide'' and ``deep and localized'' galaxy surveys, respectively.
Errors refer to our conservative bechmark model. 
Numbers in parenthesis are the marginalized errors obtained assuming perfect knowledge of the standard cosmological parameters.}
\label{tab:mg_constraints}
\end{table}

In other words, Modified Gravity effects are highly suppressed for small values of~$k_\mathrm{mg}$, therefore in those scenarios we expect to find weaker constraints on Modified Gravity parameters, as we show in table~\ref{tab:mg_constraints} both for GW alone and by cross-correlating them with galaxy surveys for our conservative astrophysical BH model.
The numbers outside parenthesis refer to the errors obtained by the full Fisher matrix, while numbers in parenthesis refer to the case where standard cosmological models are perfectly known, and we quote them to show how tighten the constraints can become when a strong prior on those parameters is imposed.\footnote{
In the case of GWs alone, fixing standard cosmological parameters help in breaking degeneracies between them and the parameters describing the Modified Gravity model and the tracer itself.
In particular, the effective bias is strongly anticorrelated with~$\omega_\mathrm{cdm}$, since both parameters controls the amplitude of the angular power spectra at large scale, hence the large improvement in the constraint.
On the other hand, in this case both~$\mu_0$ and~$\eta_0$ are almost uncorrelated with standard cosmological parameters, hence the tightening of the constraint is milder.
In the GW$\times$LSS cases, galaxy surveys provide very tight constraints on their own, thus fixing extra parameters does not improve the overall constraints on ``New Physics'' parameters significantly.
}
In particular, the difference in errors between the~$k_\mathrm{mg}=10^{-3}\ \mathrm{Mpc}^{-1}$ and the~$k_\mathrm{mg}=10^{-2}\ \mathrm{Mpc}^{-1}$ scenarios is about one order of magnitude, both for~$\mu_0$ and~$\eta_0$, while the difference between the~$k_\mathrm{mg}=10^{-2}\ \mathrm{Mpc}^{-1}$ and~$k_\mathrm{mg}=10^{-1}\ \mathrm{Mpc}^{-1}$ cases is a factor few, both with and without galaxy surveys.
Note that for large values of~$k_\mathrm{mg}$ GWs alone can provide independent and complementary constraints that are comparable to existing ones~\cite{aghanim:planckcosmoparameters, ade:planckmodifiedgravity}.
Moreover, we find that errors on the standard cosmological parameters of table~\ref{tab:cosmo_constraints} are almost unaffected by the exact choice of~$k_\mathrm{mg}$. 
In general, we conclude that SW and DL surveys will be equally competitive in constraining Modified Gravity parameters when cross-correlated to GWs, despite the latter being intrinsically better at constraining standard cosmological parameters, and they will provide constraints that are around one order of magnitude tighter than existing ones.

Moving to the nDGP model, we report errors on the~$\Omega_\mathrm{rc}$ parameter also in table~\ref{tab:mg_constraints}. 
Also in this case we notice that both galaxy surveys have the same constraining power in terms of ruling out the nDGP model. 
Moreover, we notice that errors in this case are comparable to existing ones, however in this case the constraints have been obtained considering exclusively linear scales.
In other words, our constraints can be considered more robust in the sense that does not depend on including the effect of non-linearities in the nDGP model.

We also provide in table~\ref{tab:optimistic_mg_constraints} of appendix~\ref{app:optimistic_mg_constraints} the constraints on Modified Gravity and nDGP parameters for the optimistic case corresponding to observe twice as many GW events per year, as some estimates suggest.
While improving by a factor two the estimated errors on the Modified Gravity and nDGP parameters in the GW alone case, the optimistic astrophysical BH model shows only a marginal improvement in the magnitude of the errors when galaxy surveys are included.
The general considerations regarding the importance of the Modified Gravity additional scale dependence we made for the conservative case apply also for the optimistic one.


\subsection{Constraints on black hole origin}
\label{subsec:black_hole_origin}

\begin{table}[t]
    \centerline{
    \begin{tabular}{|c|c|c|}
    \hline
    Tracers & Conservative $\sigma_{b^\mathrm{eff}_\mathrm{GW}}$ & Optimistic $\sigma_{b^\mathrm{eff}_\mathrm{GW}}$ \\
    \hline
    \hline
    SW$\times$HLVIK & $0.49$ & $0.35$ \\
    DL$\times$HLVIK & $1.36$ & $0.96$ \\
    \hline
    \end{tabular}}
    \vspace{0.3cm}
    \centerline{
    \begin{tabular}{|c|c|c|}
    \hline
    Tracers & Conservative $\sigma_{b^\mathrm{eff}_\mathrm{GW}}$ & Optimistic $\sigma_{b^\mathrm{eff}_\mathrm{GW}}$ \\
    \hline
    \hline
    SW$\times$ET2CE & $0.09$ & $0.08$ \\
    DL$\times$ET2CE & $0.08$ & $0.06$ \\
    \hline
    \end{tabular}}
\caption{Constraints on the GW effective bias parameter using second (\textit{left table}) and third generation (\textit{right table}) detector networks.
SW and DL refer to ``shallow and wide'' and ``deep and localized'' galaxy surveys, respectively.}
\label{tab:GW_bias_constraints}
\end{table}

For the purpose of estimating the SNR in equation~\eqref{eq:SNR}, we consider the astrophysical BH population as fiducial model, and the astrophysical-primordial BH population described by the parameters~$f_\mathrm{PBH}-\mathcal{A}_\mathrm{m}$ as alternative model.
We report the marginalized errors on the GW effective bias both for second and third generation detector networks in table~\ref{tab:GW_bias_constraints} for both DL and SW surveys. 
While for the second generation case a SW galaxy survey performs significantly better than a DL one, due to the wider coverage of the sky, for the third generation case the two surveys tighten the constraint up to one order of magnitude and have substantially equivalent constraining power, since now GW events can also explore the Universe at higher redshift similarly to the DL survey. 

Even though we are not interested in investigating models where Modified Gravity and PBHs describe simultaneously the dark energy and dark matter sectors, respectively, we can briefly comment on the magnitude of errors on GW effective bias in table~\ref{tab:mg_constraints} and~\ref{tab:GW_bias_constraints}.
We notice that errors on the GW effective bias appear to be minimally affected by the choice of dark energy model, whether it is a pure cosmological constant, a Modified Gravity or a nDGP model, even if there is some residual correlation that make the constraints looser when deviations from General Relativity are stronger.
Therefore, we can rather safely conclude that our prospect of disentangling the origin of BH binary progenitors are only minimally dependent on our description of the dark energy sector. 

\begin{figure}[t]
    \centerline{
    \includegraphics[width=\columnwidth]{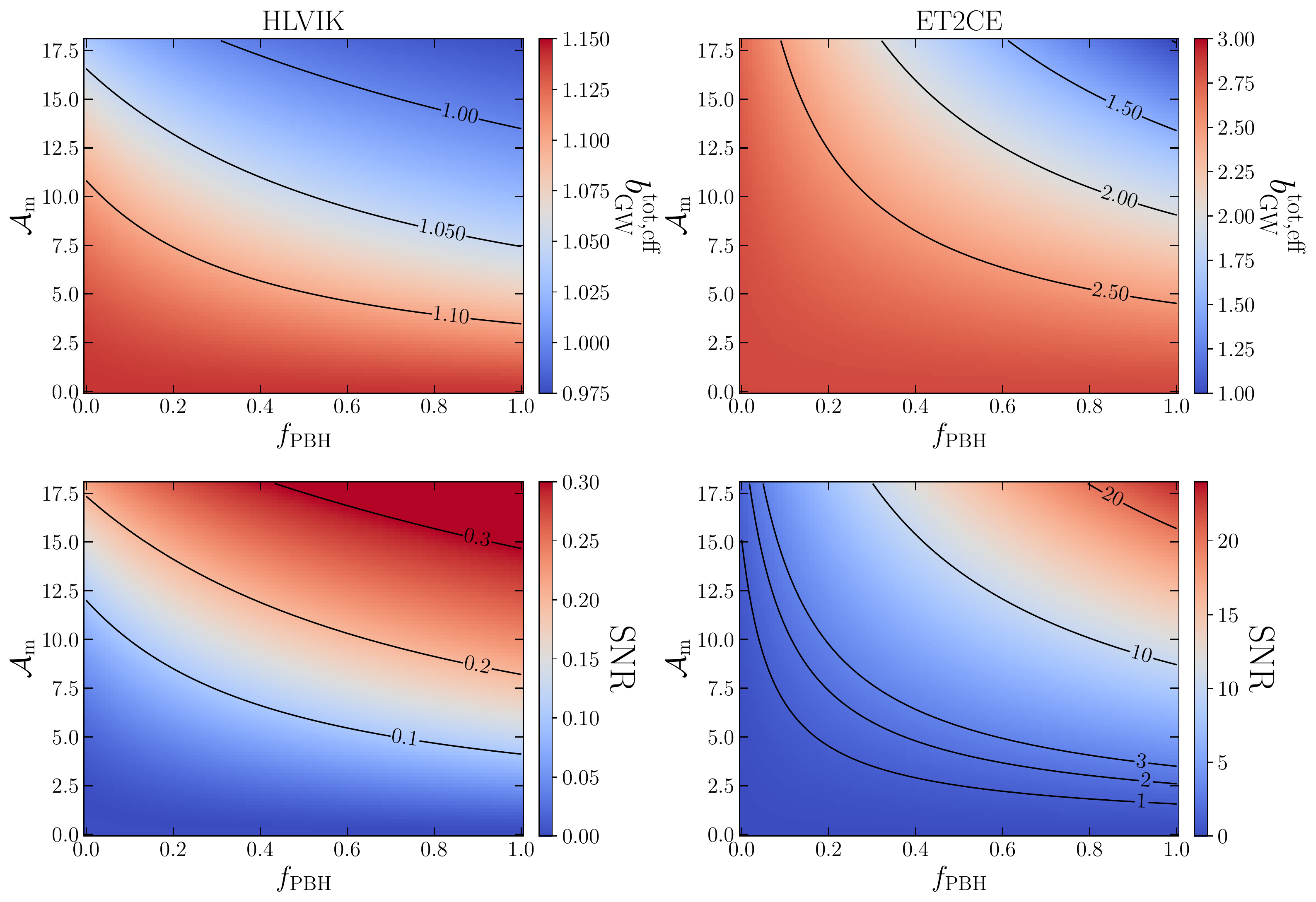}}
\caption{Total effective bias of the mixed BH binary population (\textit{top panels}) and~$\mathrm{SNR}$ obtained assuming astrophysical BHs are the fiducial model (\textit{bottom panels}), both for second (\textit{left panels}) and third generation (\textit{right panels}) GW detector network.
The fiducial model is given by the conservative astrophysical BH benchmark model.
The effective bias of the alternative model is defined in equation~\eqref{eq:mixed_model_bias_functions}.
For the HLVIK case we consider the cross-correlation with the SW survey since it is the most promising case, while for the ET2CE case the results are equivalent for both kind of galaxy surveys.}
\label{fig:SNR_ABH_PBH}
\end{figure}

In figure~\ref{fig:SNR_ABH_PBH} we show the GW effective bias for the mixed BH populations (\textit{top panels}) and the corresponding~$\mathrm{SNR}$ for second and third generation GW detector networks (\textit{bottom panels}).
The SNRs refer to the case where GWs are cross-correlated with the SW galaxy survey since in the second generation case it outperforms the DL one, while in the third they are almost equivalent.
We note that second generation GW detector networks are inefficient in discriminating the origin of the BHs because of \textit{(i)} the low number of detected events, which implies a more noisy measurements of the GW clustering statistics, and \textit{(ii)} the fact that detected sources are mainly at low redshift, where relative differences between the models are less accentuated, as the interested reader can see in figure~\ref{fig:BH_2G_mixed_models} of appendix~\ref{app:mixed_BBH_2G}.
On the other hand, the third generation case is radically different.
In particular, we will be able to statistically infer the presence of a primordial component of sources at more than~$3\sigma$ confidence in a large portion of the~$f_\mathrm{PBH}-\mathcal{A}_\mathrm{m}$.
Most notably, if PBHs are the dark matter, i.e., if~$f_\mathrm{PBH}=1$, we will be able to assess it at the~$20\sigma$ level, making GW-galaxy cross-correlation one of the most important probes to determine PBH existence in the~$\mathcal{O}(10)\ M_\odot$ mass range.
If we consider an optimistic astrophysical BH model with twice the number of events of the conservative one, the improvement in the SNR in the second generation detector network case is of the order of~$30\%$, hence it is still insufficient in effectively discriminating the nature of the compact objects.
On the other hand, the same improvement in the third generation detector network case further expands the parameter space that can be effectively probed by the cross-correlation technique.

In parallel to this work, another article investigated PBH clustering~\cite{libanore:pbhclustering}. 
The two approaches are comparable yet different, since we focus our attention to produce realistic catalogs of detected events and on the existence of an intrinsic correlation between different PBH binary formation channels given by the abundance parameter~$f_\mathrm{PBH}$. The treatment of astrophysical and binary PBH formation uncertainties is also somewhat different but conservative in both cases.
Nevertheless, constraints are comparable once we rescale appropriately the early time PBH binary local merger rate and the PBH abundance.


\section{Conclusions}
\label{sec:conclusions}

The emerging field of GW astronomy offers outstanding opportunities to address multiple open questions on the nature of our Universe. 
Future GW detectors, thanks to their increased sensitivity, are expected to observe tens of thousand of GW events per year, shedding light on the physics behind GW sources. 
Along with this unprecedented opportunity to explore the Universe through GW radiation, the Cosmology community has to face a new challenge: developing new techniques and ideas to fully maximize the scientific outcome of this new kind of dataset.
In this work we explore this avenue by investigating whether GW clustering alone or in combination with galaxies can improve our understanding of both the dark matter and dark energy sectors.
In fact GWs, since they are emitted by sources which live in a variety of environments (galaxies, dark matter halos, and so on), represent a legitimate tracers of the large-scale structure of the Universe. 

Since a priori it is not known which kind of galaxy survey is more suitable to be correlated with GWs, in this work we consider both a shallow and wide (SW) and a deep and localized (DL) galaxy surveys.
Moreover, since the intrinsic origin of BHs is currently unknown, we analyze both astrophysical and primordial BHs as potential sources for detected GW events.
We created realistic catalogs of GW events following specific state-of-the-art binary models and we determine the clustering properties of GW events detected by second and third generation detector networks using the external modules of~\texttt{CLASS\_GWB}.
We generalized the public code~\texttt{Multi\_CLASS} to include extension of General Relativity described by the standard~$\mu-\eta$ parametrization, and we use it to compute the GW and galaxy clustering statistics. 

Using astrophysical BHs alone, we could constrain extensions of General Relativity approximately at the same level of existing constraints using exclusively linear scales.
In other words, GWs represent a tracer of the large-scale structure of the Universe that is independent from, for instance, galaxies, and that can be used to place constraints that are subject to a different set of systematic errors. 
Moreover, these constraints do not depend on the modelling of non-linear scales, which is not a well explored territory in Modified Gravity theories.
By including cross-correlations with galaxies we find competitive constraints even for our conservative astrophysical benchmark model.
We also showed how the strength of the constraints explicitly depend on the presence of extra scales in the Modified Gravity theories, providing a fair representation (which is sometimes lacking in the literature) of the constraining power of this new probe.

Regarding the possibility of assessing whether merging binary BHs might have astrophysical or primordial origin, we explored two popular PBH binary formation channels and we compared them to the standard astrophysical one.
Even though the number of events detected by a second generation GW detector network at design sensitivity appear to be insufficient to establish the origin of the binary due to volume selection effects that smears out the relative differences between models, we find that third generation detectors will be able to either rule in or out PBHs as a major component of the dark matter at more than~$3\sigma$ confidence level in the tens of solar masses range.
Therefore GW clustering can be effectively used to robustly constraint the abundance of PBH in a mass range that is heavily probed~\cite{zumalacarregui:pbhconstraints, brandt:pbhconstraints, koushiappas:pbhconstraints, zoutendijk:pbhconstraints, afshordi:pbhconstraints, murgia:pbhconstraints, monroyrodriguez:pbhconstraints, alihaimoud:pbhconstraints, poulin:pbhconstraints, serpico:pbhconstraints, bernal:pbhconstraints, gaggero:accretionconstraints, manshanden:accretionconstraint, hektor:accretionconstraintII, hektor:accretionconstraint, hutsi:accretionconstraint, mena:accretionconstraint, piga:accretionconstraints}, but still suffers from large theoretical uncertainties due to the theoretical modelling of the observables and the unknown PBH mass distribution, see, e.g., refs.~\cite{bellomo:pbhconstraints, piga:accretionconstraints}.
Finally, we note that it is more difficult to probe PBH abundances of~$f_\mathrm{PBH}\lesssim 0.1$, independently on the relative abundance of late time and early time PBH binaries.

In this work we considered only dark sirens as a GW tracer to provide a conservative result.
However other types of merging events, such as BH-neutron star or neutron star-neutron star mergers, can be reasonably included in this kind of analysis.
On top of a straightforward increase in the number of objects used to create GW maps of the sky, which certainly helps in diminishing the noisiness of the maps, in certain instances these kind of events might have a electromagnetic counterpart, which also allows for a better localization.
However, when comparing the total number of events detected both by second and third generation GW detector networks, we still find that the number of detected binary BH mergers dominates over BH-neutron star and neutron star-neutron star events.
Therefore, even in the (very) optimistic case where every single event involving a neutron star has an electromagnetic counterpart, we do not expect the improvement to be substantial for the kind of analysis performed here since the resolution of the maps would still be limited by binary BH events.

Given the potential of the cross-correlation technique to test beyond General Relativity models, it would be interesting to perform this kind of analysis using a Modified Gravity model for which we know its strong field regime predictions, and to account in a self-consistent way for all aspects of these theories ranging from differences in the emitted GW waveform to differences in the GW propagation in a perturbed Universe.
Further investigations on different GW sources also appear to be a promising avenue, including in different frequency bands such as the mHz (probed by LISA) or the nHz (probed by PTA), especially when the properties of these sources heavily rely on properties of their environment.


\acknowledgments

MB thanks Andrea Begnoni and Giulio Scelfo for interesting discussions. The authors thank Sarah Libanore for useful discussions.
NB acknowledges partial support from the National Science Foundation (NSF) under Grant No.~PHY-2112884.
AR acknowledges funding from the Italian Ministry of University and Research (MIUR) through the ``Dipartimenti di eccellenza'' project ``Science of the Universe''.


\appendix
\section{Number count contributions}
\label{app:numbercountfluctuation_transferfunction}

The contributions to $\Delta^X_\ell(k,z)$ introduced in equation~\eqref{eq:numbercountfluctuation_projectioneffects} read as
\begin{equation}
    \begin{aligned}
        \Delta^{X,\mathrm{den}}_\ell(k,z) &= b_X j_\ell D(k, \tau_z), \\
        \Delta^{X,\mathrm{vel}}_\ell(k,z) &= \frac{k}{\mathcal{H}}\partial^2_y j_\ell V(k, \tau_z) + \left[\left(f^\mathrm{evo}_X-3\right)\frac{\mathcal{H}}{k}j_\ell + \left(\frac{\mathcal{H}'}{\mathcal{H}^2}+\frac{2-5s_X}{r(z)\mathcal{H}}+5s_X - f^\mathrm{evo}_X\right)\partial_y j_\ell\right] V(k, \tau_z), \\
        \Delta^{X,\mathrm{len}}_\ell(k,z) &= \ell(\ell+1)\frac{2-5s_X}{2} \int^{r(z)}_0 dr \frac{r(z)-r}{r(z)r} \left[\Phi(k,\tau_z)+\Psi(k,\tau_z)\right]j_\ell(kr), \\
        \Delta^{X,\mathrm{gr}}_\ell(k,z) &= \left[\left(\frac{\mathcal{H}'}{\mathcal{H}^2} + \frac{2-5s_X}{r(z)\mathcal{H}}+5s_X-f^\mathrm{evo}_X+1\right) \Psi(k,\tau_z) + (5s_X-2) \Phi(k,\tau_z) + \mathcal{H}^{-1}\Phi'(k,\tau_z)\right] j_\ell \\
        &\qquad + \int^{r(z)}_0 dr \frac{2-5s_X}{r(z)}\left[\Phi(k,\tau)+\Phi(k,\tau)\right]j_\ell(kr) \\
        &\qquad + \int^{r(z)}_0 dr\left(\frac{\mathcal{H}'}{\mathcal{H}^2}+\frac{2-5s_X}{r(z)\mathcal{H}}+5s_X-f^\mathrm{evo}_X\right)_{r(z)} \left[\Phi'(k,\tau)+\Psi'(k,\tau)\right]j_\ell(kr).
    \end{aligned}
\end{equation}

As we already mentioned, $b_X$ is the bias parameter of the tracer~$X$, $s_X$ is the magnification bias parameter and $f^\mathrm{evo}_X$ is the evolution bias parameter. 
Then, $r(z)$ is the radial comoving distance at redshift~$z$, $\tau_z = \tau_0-r(z)$ and $\tau = \tau_0-r$ are conformal times of perturbations that we observe at the comoving distances $r(z)$ and $r$, with $\tau_0$ being the conformal age of the Universe. 
Besides, $\mathcal{H}=a'/a$ is the Hubble expansion rate and the prime $'$ denotes a derivative with respect to conformal time, $D$ is the gauge invariant density perturbation, $V$ is the gauge invariant velocity perturbation, $\Phi$ and $\Psi$ are Bardeen potentials. 
Finally, $j_\ell=j_\ell(kr(z))$ are Bessel functions and $\partial^n_y = d^n/d(kr(z))^n$.


\section{Implementation of Modified Gravity models in~\texttt{Multi\_CLASS}}
\label{app:MG_Multi_CLASS}

The implementation of the Modified Gravity and nDGP models in~\texttt{Multi\_CLASS} follows the approach described in ref.~\cite{baker:modifiedgravity}. 
In particular, we modify the evolution equations for the metric potentials, which in the~$\mu-\eta$ parametrization now read as
\begin{equation}
    \begin{aligned}
        \Psi &= \frac{\Phi}{\eta} - 12\pi G\left(\frac{a}{k}\right)^2 (\bar{\rho}_\mathrm{tot}+\bar{p}_\mathrm{tot})\sigma_\mathrm{tot}, \\
        \displaystyle \Phi' &= \frac{\left[\Phi\left(\frac{\mu'}{\mu} + \frac{\eta'}{\eta} - \mathcal{H}\right) + \frac{9}{2}\frac{\mathcal{H}^2}{k^2} \Omega_\mathrm{m} \theta_\mathrm{m} \mu \eta \left(\frac{1}{3}+\frac{\mathcal{H}^2-\mathcal{H}'}{k^2}\right) - \frac{9}{2}\frac{\mathcal{H}^2}{k^2} \Omega_\mathrm{m} \mu \eta \Psi \mathcal{H}\right]}{\left(1 + \frac{9}{2}\frac{\mathcal{H}^2}{k^2} \Omega_\mathrm{m} \mu \eta\right)}
    \end{aligned}
\end{equation}
where~$\bar{\rho}_\mathrm{tot}$ and~$\bar{p}_\mathrm{tot}$ are the total background energy density and pressure, respectively, $\sigma_\mathrm{tot}$ is the total anisotropic stress, $\Omega_\mathrm{m}(z)$ is the fractional abundance of matter at redshift~$z$ and~$\theta_\mathrm{m}$ is the matter velocity divergence. 
The~$\Lambda$CDM limit is recovered for constant~$\mu=\eta=1$, as discussed in ref.~\cite{baker:modifiedgravity}.

It is trivial to take derivatives of the~$\mu-\eta$ functions in equations~\eqref{eq:Modified_Gravity_mu_eta} in our parametrization of the Modified Gravity scenario, therefore we do not report them here.
On the other hand, in the nDGP model, we have
\begin{equation}
    \mu' = -\frac{\beta'}{3\beta^2}, \qquad \eta' = \frac{6\beta'}{(3\beta+1)^2}.
\end{equation}
Deriving equation~\eqref{eq:nDGP_beta} with respect to the conformal time we obtain
\begin{equation}
    \beta' = \frac{H'}{H_0\sqrt{\Omega_\mathrm{rc}}} + \frac{1}{3H_0\sqrt{\Omega_\mathrm{rc}}} \left(\frac{H'}{aH} \right)' = \frac{1}{3H_0\sqrt{\Omega_\mathrm{rc}}} \left(\frac{H''}{aH} + 2H' - \frac{(H')^2}{aH^2} \right),
\end{equation}
where the explicit expression for~$H''$ can be calculated from the Friedmann equation
\begin{equation}
    2\dot{H} + 3H^2 = -8\pi G \bar{p}_\mathrm{tot}, 
\end{equation}
finding that
\begin{equation}
     H'' + 2aHH' = -4\pi G a \bar{p}_\mathrm{tot}' = \frac{16\pi G}{3}a^2H \bar{\rho}_\mathrm{r} = 2 a^2H^3 \frac{\bar{\rho}_\mathrm{r}}{\bar{\rho}_\mathrm{tot}},
\label{eq:Hprimeprime}
\end{equation}
where~$\bar{\rho}_\mathrm{r}$ is the radiation background energy. 
More in detail, in equation~\eqref{eq:Hprimeprime} we use
\begin{equation}
    \bar{p}_\mathrm{tot} = \bar{p}_\Lambda + \bar{p}_\mathrm{m} + \bar{p}_\mathrm{r} \qquad \Longrightarrow \qquad  a\bar{p}_\mathrm{tot}' = a^3H\frac{d\bar{p}_\mathrm{tot}}{da} = - \frac{4a^2H\bar{\rho}_\mathrm{r}}{3},
\end{equation}
with~$\bar{p}_\Lambda$ and~$\bar{p}_\mathrm{m}$ being the cosmological constant and matter background pressure, respectively. 
The RHS of equation~\eqref{eq:Hprimeprime} is expected to be very suppressed during the matter and dark energy dominated eras.


\section{Gravitational wave detector networks}
\label{app:detector_networks}

\begin{table}[t]
\centerline{
\begin{tabular}{|c|c|c|c|c|c|c|}
\hline
Detector        & $\beta$ & $\lambda$ & $\varphi_1$ & $\omega_1$ & $\varphi_2$ & $\omega_2$ \\
\hline
\hline
LIGO Hanford    & $0.8108$ &  $-2.0841$ & $2.1991$ & $-6.195\times 10^{-4}$ & $3.7699$ & $1.25\times 10^{-5}$ \\
\hline
LIGO Livingston & $0.5334$ & $-1.5843$ & $3.4508$ & $-3.121\times 10^{-4}$ & $5.0216$ & $-6.107\times 10^{-4}$ \\
\hline
Virgo           & $0.7615$ & $0.1833$ & $1.2316$ & $0.0$ & $2.8024$ & $0.0$ \\
\hline
KAGRA           & $0.6355$ & $2.3965$ & $0.4939$ & $3.141\times 10^{-3}$ & $2.0647$ & $-3.627\times 10^{-3}$ \\
\hline
LIGO India      & $0.3423$ & $1.3444$ & $4.2304$ & $0.0$ & $5.8012$ & $0.0$ \\
\hline
\end{tabular}}

\vspace{0.5cm}

\centerline{
\begin{tabular}{|c|c|c|c|c|c|c|}
\hline
Detector        & $\beta$ & $\lambda$ & $\varphi_1$ & $\omega_1$ & $\varphi_2$ & $\omega_2$ \\
\hline
\hline
ET1 & $0.7615$ & $0.1833$ & $1.2316$ & $0.0$ & $2.2788$ & $0.0$ \\
\hline
ET2 & $0.7630$ & $0.1841$ & $3.3260$ & $0.0$ & $4.3732$ & $0.0$ \\
\hline
ET3 & $0.7627$ & $0.1819$ & $5.4204$ & $0.0$ & $0.1844$ & $0.0$ \\
\hline
CE1 & $0.7649$ & $-1.9692$ & $0.0$ & $0.0$ & $1.5708$ & $0.0$ \\
\hline
CE2 & $0.5788$ & $-1.8584$ & $2.6180$ & $0.0$ & $4.1888$ & $0.0$ \\
\hline
\end{tabular}}
\caption{Detector latitude~$\beta$, longitude~$\lambda$ and arm orientation angles~$(\varphi_1,\omega_1,\varphi_2,\omega_2)$ in radians for the second (\textit{upper table}) and third (\textit{lower table}) generation GW detector network. 
We choose Einstein Telescope location to be at the Germany-Belgium-The Netherlands border, while for Cosmic Explorer we consider two sites on the USA mainland already used for preparatory studies~\cite{borhanian:cepotentiallocation, gossan:cepotentiallocation}. 
Notation follows that of ref.~\cite{bellomo:classgwb}.}
\label{tab:detector_constants}
\end{table}

In this work we consider both second and third generation GW observatories to compare their performances. 
Regarding the former, we consider a network made by LIGO Hanford and Livingston, Virgo, KAGRA and LIGO India L-shaped interferometers. 
On the other hand, for the latter, we choose a network made by Einstein Telescope (composed by three V-shaped interferometer) and Cosmic Explorer (two L-shaped interferometer, with arm length of~$40$ and~$20$ km for the main and secondary site, respectively). 
Existing and future detector location (latitude and longitude) and orientation angles of the arms are reported in table~\ref{tab:detector_constants}.

\begin{figure}[t]
\centerline{
\includegraphics[width=\columnwidth]{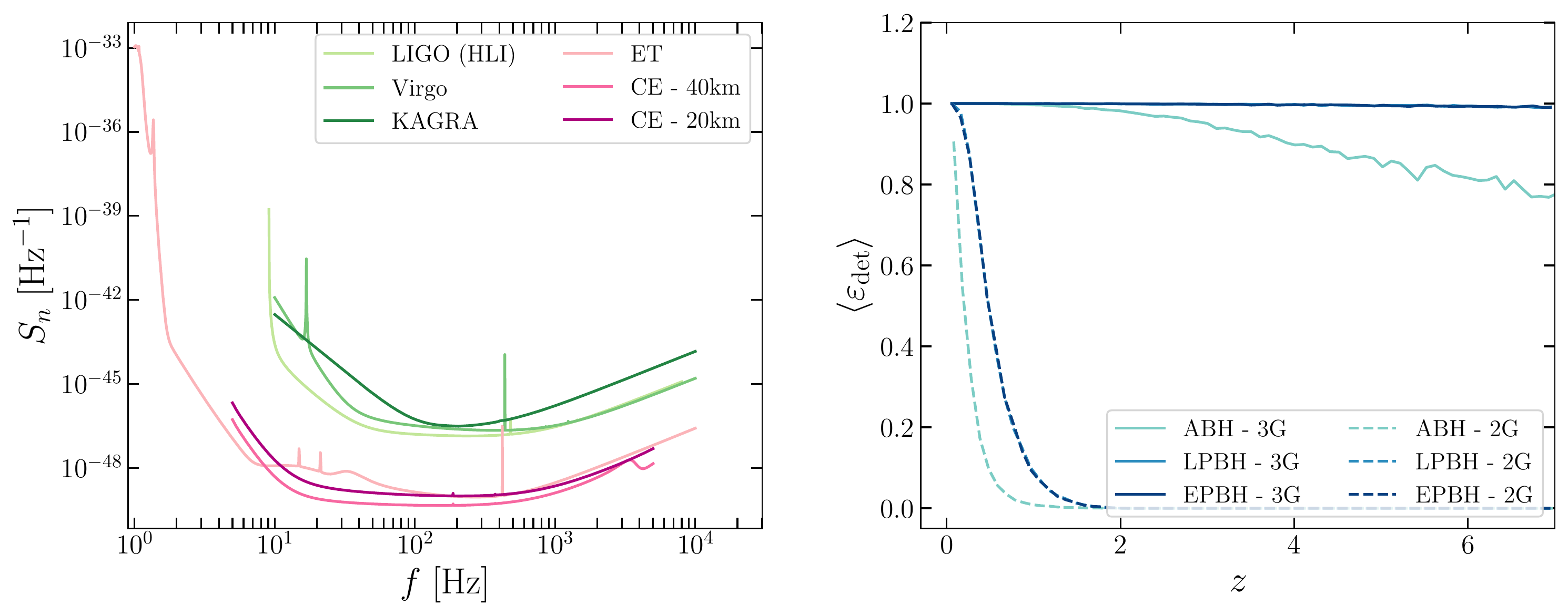}}
\caption{\textit{Left panel}: One-sided noise spectral density for individual detectors belonging to second and third generation detector networks. 
\textit{Right panel}: Average detection efficiencies for different BBH sources and detector networks.}
\label{fig:Sn_epsilon_detectors}
\end{figure}

Regarding ET, we consider the ``ET-D'' base sensitivity for the one-sided noise spectral density~$S_n$. 
The nominal curve is given for a~$10\ \mathrm{km}$ L-shaped interferometer hence it has to be rescaled: following the ET community guidelines we use~$S^\mathrm{real}_n = \left(2\sqrt{2}/3\right)^2 S^\mathrm{nominal}_n$. 
The frequency range considered for the three detectors is~$\left[1,10000\right]\ \mathrm{Hz}$. 
Noise is assumed to be uncorrelated in the three interferometers. 
Regarding CE, we consider two L-shaped interferometers, one with 40-km arms, the other with 20km arms. 
The comparison between second and third generation noise spectral densities is presented on the left panel of figure~\ref{fig:Sn_epsilon_detectors}.
We choose~$\rho_\mathrm{det}=12$ as SNR threshold for both second and third generation detector networks.
We show on the right panel of figure~\ref{fig:Sn_epsilon_detectors} the average detection efficiency as a function of redshift for both detector network and for all binary populations. 
For each binary population the displayed average efficiency is an average of the detection efficiency of five different catalogs of~$10^5$ events.
The noisy behaviour of the curves at high redshift is due to the low number of events in that epoch.

\begin{figure}[ht]
\centerline{
\includegraphics[width=\columnwidth]{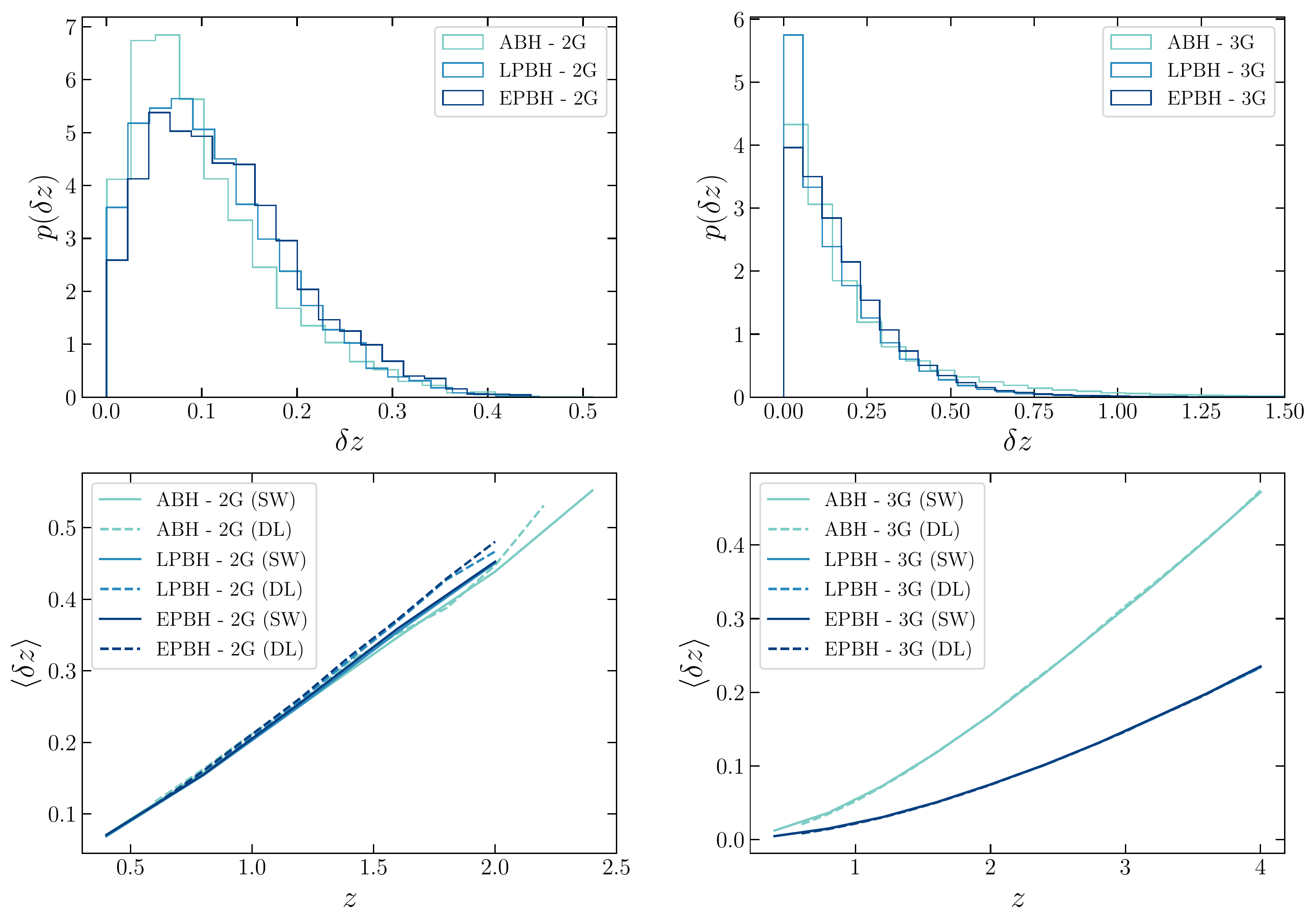}}
\caption{\textit{Top panels}: redshift error probability distribution function for second (\textit{left panel}) and third (\textit{right panel}) generation GW detector network for all binary populations considered in this work.
\textit{Bottom panels}: average redshift localization error in different redshift bins for second (\textit{left panel}) and third (\textit{right panel}) generation GW detector network for all binary populations considered in this work. Solid and dashed lines refer to the redshift binning chosen for the SW and DL surveys, respectively.
For comparison, the width of redshift bins in our analysis is~$2\Delta z=0.4$ and~$2\Delta z=0.2$ for SW and DL surveys, respectively.}
\label{fig:deltaz_pdf}
\end{figure}

Given the catalogs of detected events for different populations, we can estimate what are the typical errors in determining the redshift of the source.
We show in top panels of figure~\ref{fig:deltaz_pdf} the redshift error probability distribution function for different detector networks and binary populations.
The errors are estimated for each event using~$\delta z_\mathrm{event}=3z_\mathrm{event}/\rho_\mathrm{event}$~\cite{calore:gwxlss, delpozzo:reshifterror} and should be compared to the redshift bin width~$2\Delta z=0.2$ and~$2\Delta z=0.4$ we use for DL and SW galaxy surveys, respectively.
We find that the median redshift error is~$\delta z \simeq 0.1$ for all populations and GW detector networks, making our binning choice broadly compatible with typical error on the source distance.
Moreover we explicitly check the magnitude of the average localization error~$\left\langle \delta z \right\rangle$ in each redshift bin, and we show it on the bottom panels of figure~\ref{fig:deltaz_pdf}.
These figures further support our binning choice, and they also show the redshift-dependent dispersion~$\sigma_z=\left\langle \delta z \right\rangle$ we later use in the computation of the GW noise in equation~\eqref{eq:gw_total_noise}.
While for second generation networks the three classes of binaries have similar~$\rho^2$ distributions, thus a similar~$\left\langle \delta z \right\rangle$ value.
On the other hand, in the case of third generation detectors, our model of PBH describes binaries that are detected with a higher signal-to-noise ratio compared to astrophysical ones, hence the better localization error in redshift.
Finally, for third generation detectors we find a high~$\delta z$ tail due to events located at high redshift, suggesting that the optimal binning strategy might be one that has narrow bins at low redshift and wide bins at high redshift, however we leave the investigating of an optimal binning strategy for future work.

Regarding the determination of the average and maximum resolution, we use the results in figure~3 of ref.~\cite{calore:gwxlss}. 
In particular, we find that on average the number of GW detected by second and third generation detector network peaks at redshift~$z^\mathrm{2G}_\mathrm{peak}\simeq 0.3, 0.4, 0.5$ and~$z^\mathrm{3G}_\mathrm{peak}\simeq 1.4, 2.2, 3.4$ for astrophysical binaries, late-time and early-time primordial binaries, respectively. 
Therefore the average resolution for the two classes of detectors is given by~$\theta^\mathrm{avg,2G}_\mathrm{res} \simeq 2.6^\circ, 3.2^\circ, 3.4^\circ$ and~$\theta^\mathrm{avg,3G}_\mathrm{res} \simeq 0.8^\circ, 1.2^\circ, 1.8^\circ$. 
The maximum resolution is given by~$\theta^\mathrm{max,2G}_\mathrm{res} = 0.6^\circ, 0.8^\circ, 1.0^\circ$ and~$\theta^\mathrm{max,3G}_\mathrm{res} = 0.1^\circ, 0.2^\circ, 0.4^\circ$, which would correspond to~$\ell^\mathrm{2G}_\mathrm{max}=300, 225, 180$ and~$\ell^\mathrm{3G}_\mathrm{max}=1200, 900, 450$; however we consider as maximum multipole~$\ell_\mathrm{max} = 200$ to avoid modelling the non-linear regime, see, e.g., ref.~\cite{bellomo:multiclass}. 
We note that these estimates are consistent also with more recent ones, see, e.g., ref.~\cite{iacovelli:gw3genforecast}.


\section{Late time primordial black hole binary formation}
\label{app:lpbh_binary_formation}

In total generality, any PBH population is described by the fractional abundance function~\cite{bellomo:pbhconstraints}
\begin{equation}
    \frac{df_\mathrm{PBH}}{dM_\mathrm{PBH}} = f_\mathrm{PBH} \frac{d\Phi_\mathrm{PBH}}{dM_\mathrm{PBH}},
\end{equation}
where the abundance parameter~$f_\mathrm{PBH} = \bar{\rho}_\mathrm{PBH} / \bar{\rho}_\mathrm{dm}$ describes the fraction of dark matter in form of PBHs, and~$d\Phi_\mathrm{PBH}/dM_\mathrm{PBH}$ describes the shape of the PBH mass distribution.
The latter function is normalized to unity by construction. 
For the rest of this appendix we consider two benchmark cases: a monochromatic mass distribution (MMD) where all compact object have mass~$M^\star_\mathrm{PBH}$
\begin{equation}
    \frac{d\Phi_\mathrm{PBH}}{dM_\mathrm{PBH}} = \delta^D(M_\mathrm{PBH}-M^\star_\mathrm{PBH}),
\end{equation}
where~$M^\star_\mathrm{PBH}=30\ M_\odot$ is our benchmark value, and a \textit{lognormal} extended mass distribution~\cite{bellomo:pbhconstraints} (EMD)
\begin{equation}
    \frac{d\Phi_\mathrm{PBH}}{dM_\mathrm{PBH}} = \frac{e^{-\frac{\log^2(M_\mathrm{PBH}/\mu)}{2\sigma^2}}}{\sqrt{2\pi}\sigma M_\mathrm{PBH}},
\end{equation}
characterized by two parameters, mean~$\mu$ and standard deviation~$\sigma$, with~$(\mu,\sigma)=(40,0.5)$ being our benchmark value.
With this choice of values we have an EMD peaked around~$30\ M_\odot$, mimicking a MMD but with some width.
Additionally, we assume PBH to be spinless, i.e., $S_\mathrm{PBH}=0$, hence the (dimensionless) spin parameter reads as $\chi_\mathrm{PBH} = cS_\mathrm{PBH}/GM^2_\mathrm{PBH} = 0$.
This assumption is consistent with the predictions of the popular models where PBH form from the collapse of large overdensities during the radiation-dominated era~\cite{takeshi:pbhspin, mirbabayi:pbhspin, deluca:pbhspin}.

Independently from the origin of the BHs, binaries are described by the total and reduced masses
\begin{equation}
    M = M_1 + M_2, \qquad \mu = \frac{M_1 M_2}{M_1 + M_2},
\end{equation}
respectively, where~$M_1, M_2$ are the two compact object masses.


\subsection{Late time binaries}

The binary formation rate density appearing in equation~\eqref{eq:late_time_merger_rate_density} is given by~\cite{bird:pbhlatetimebinaries, clesse:pbhlatetimebinaries}
\begin{equation}
    \bar{R}_\mathrm{bf} = \int_{M_{h}^\mathrm{min}}^{M_{h}^\mathrm{max}} dM_h \frac{dn_h}{dM_h} \bar{R}_{\mathrm{bf},h}(M_h,z),
\label{eq:binary_formation_rate_density}
\end{equation}
where~$M_h$ is the dark matter halo mass, $dn_h/dM_h$ is the halo number density given in ref.~\cite{puebla:halonumberdensity} and~$\bar{R}_{\mathrm{bf},h}$ is the binary formation rate per halo
\begin{equation}
    \bar{R}_{\mathrm{bf},h}(M_h, z) = f^2_\mathrm{PBH} \int d^3r dM_1 dM_2 \rho_h^2(r) \frac{d\Phi_\mathrm{PBH}}{dM_1}\frac{d\Phi_\mathrm{PBH}}{dM_2} \frac{\left\langle \sigma_\mathrm{dc}v_\mathrm{rel} \right\rangle}{2 M_1 M_2},
\label{eq:late_time_bfr_per_halo}
\end{equation}
with~$\left\langle \sigma_\mathrm{dc}v_\mathrm{rel} \right\rangle$ is the velocity-averaged direct capture cross section. 
Accordingly with ref.~\cite{bird:pbhlatetimebinaries}, we consider~$M^\mathrm{min}_h=400\ M_\odot$ as minimum halo mass and, as maximum halo mass, the redshift-dependent maximum mass~$M^\mathrm{max}_h(z)$ a virial structure can have at redshift~$z$~\cite{behroozi:universemachine}, even though the latter does not influence the total merger rate because it is dominated by low mass halos.


\subsubsection{Dark matter halos}

Assuming that dark matter halos made of PBH have a similar phenomenology to halo in cold matter scenarios,\footnote{Note that other choices of dark matter density profile (not divergent at the origin) proved to not influence significantly the final PBH biary formation rate~\cite{bird:pbhlatetimebinaries}.} we describe the spatial distribution of matter inside dark matter halos by a Navarro-Frenk-White density profile~\cite{navarro:nfwprofile}
\begin{equation}
    \rho_h(r) = \frac{\rho_s}{\frac{r}{r_s}\left(1+\frac{r}{r_s}\right)^2},
\end{equation}
where~$\rho_s$ and~$r_s$ are the characteristic density and radius. 
Halos are commonly described by virial radius~$r_h$, average density~$\rho_c \Delta_c$, where~$\rho_c = 3H^2(z)/8\pi G$ is the critical density of the Universe and~$\Delta_c=200$ is the most common choice for the virial overdensity threshold, and mass
\begin{equation}
    M_h = 4\pi \int_0^{r_h} dr r^2 \rho_h(r) = 4\pi r_s^3 \rho_s g(C) = \frac{4\pi}{3} r_h^3 \rho_c \Delta_c,
\label{eq:halo_mass_definitions}
\end{equation}
where we define the concentration parameter~$C=r_h/r_s$ and the function
\begin{equation}
    g(C) = \log(1+C) - \frac{C}{1+C}.
\end{equation}
From equation~\eqref{eq:halo_mass_definitions} we have that the characteristic density and radius read in terms of the concentration parameter as
\begin{equation}
    \rho_s = \frac{\Delta_c H^2 C^3}{8\pi G g(C)}, \qquad r_s = \left( \frac{2GM_h}{\Delta_c H^2 C^3} \right)^{1/3},
\end{equation}
therefore the integration over the halo profile in equation~\eqref{eq:late_time_bfr_per_halo} factorizes out and is given by
\begin{equation}
    \int d^3r \rho_h^2(r) = \frac{\Delta_c M_h H^2 C^3}{24\pi G g^2(C)} \left[1 - (1+C)^{-3} \right].
\end{equation}

The concentration parameter is a function both of halo mass and redshift, and its value can be inferred and fitted from numerical simulations~\cite{prada:concentrationparameter, ludlow:concentrationparameter}. In this work we use the fitting formulas of ref.~\cite{ludlow:concentrationparameter}, 
 which reads as
\begin{equation}
    C(M,z) = c_0 \left(\frac{\nu}{\nu_0}\right)^{-\gamma_1} \left[ 1 + \left(\frac{\nu}{\nu_0}\right)^{1/\beta} \right]^{\beta(\gamma_1 - \gamma_2)},
\end{equation}
where~$\nu = \delta_c / \sigma(M,z)$ is the ``peak height'', $\delta_c = 1.686$ is the linearly extrapolated critical overdensity for spherical collapse in Einstein-de Sitter, $\sigma$ is the variance of the matter field, the fitting parameters are given by
\begin{equation}
    \begin{aligned}
        c_0 &= 3.395 (1 + z)^{-0.215}, \qquad \beta = 0.307 (1 + z)^{0.540}, \\
        \gamma_1 &= 0.628 (1 + z)^{-0.047}, \qquad \gamma_2 = 0.317 (1 + z)^{-0.893}, \\
        \nu_0 &= \left[4.135 - 0.564 (1 + z) - 0.210(1 + z)^2 + 0.0557(1 + z)^3 - 0.00348(1 + z)^4\right] / D(z), \\
    \end{aligned}
\end{equation}
and~$D(z)$ is the linear growth factor. 
For the matter variance and linear growth factor we use the fitting formulas of refs.~\cite{puebla:halonumberdensity} and~\cite{lahav:lineargrowthfactor, carroll:lineargrowthfactor}, respectively (see also appendix D of ref.~\cite{bellomo:classgwb}).

Regarding the distribution of velocities inside dark matter halos, it is helpful to define two quantities starting from the expression for the circular velocity~$v_c=GM(r)/r$ for a halo in virial equilibrium: $v_h$ and~$v_\mathrm{dm}$. The former is called ``virial'' velocity and it is the circular velocity at the edge of the halo, i.e.,
\begin{equation}
    v_h = v_c(r_h) = \left(\frac{\Delta_c}{2}\right)^{1/6} (GM_hH)^{1/3},
\end{equation}
while the latter is defined as the escape velocity at radius~$r_\mathrm{max}=C_\mathrm{max} r_s$, with~$C_\mathrm{max}=2.1626$, which represents the position of the maximum of the circular velocity, and reads as
\begin{equation}
    v_\mathrm{dm} = \sqrt{2} v_c(r_\mathrm{max}) = \sqrt{2 \frac{C}{C_\mathrm{max}} \frac{g(C_\mathrm{max})}{g(C)}} v_h.
\end{equation}
Numerical N-body simulations suggest that the relative velocity~$v_\mathrm{rel}$ probability distribution function of pair of objects within a dark matter halo can be described by a truncated Maxwell-Boltzmann distribution, i.e., as~\cite{bird:pbhlatetimebinaries, mao:relativevelocities, cholis:orbitaleccentricity}
\begin{equation}
    p(v_\mathrm{rel}) \propto e^{-v^2_\mathrm{rel}/v^2_\mathrm{dm}} - e^{-v^2_h/v^2_\mathrm{dm}}.
\label{eq:relative_velocity_pdf}
\end{equation}


\subsubsection{Direct capture process}

The direct capture cross section reads as~\cite{quinlan:directcapturecrosssection}
\begin{equation}
    \sigma_\mathrm{dc} = 2\pi \left( \frac{85\pi}{6\sqrt{2}} \right)^{2/7} \frac{G^2 M^{12/7}\mu^{2/7}}{c^{10/7}v^{18/7}_\mathrm{rel}}.
\end{equation}
Since the cross section strongly depends on the inverse of the relative velocity, the process will be more effective in low-mass dark matter halos. 
Using equation~\eqref{eq:relative_velocity_pdf} we find the velocity-averaged cross section
\begin{equation}
    \begin{aligned}
        \left\langle \sigma_\mathrm{dc} v_\mathrm{rel} \right\rangle &= \int d^3 v_\mathrm{rel} \sigma_\mathrm{dc} v_\mathrm{rel} p(v_\mathrm{rel}) = 2\pi \left( \frac{85\pi}{6\sqrt{2}} \right)^{2/7} \frac{G^2 M^{12/7}\mu^{2/7}}{c^{10/7}} \int_0^{v_h} d v_\mathrm{rel} v^{3/7}_\mathrm{rel} p(v_\mathrm{rel}) \\
        &= \frac{12\pi}{5} \left( \frac{85\pi}{6\sqrt{2}} \right)^{2/7} \frac{G^2 M^{12/7}\mu^{2/7}}{c^{10/7}v^{11/7}_\mathrm{dm}} \frac{5\Gamma[5/7,0] - 5\Gamma[5/7,V^2] - 7V^{10/7}e^{-V^2}}{3\sqrt{\pi}\mathrm{Erf}[V] - (6V+4V^3)e^{-V^2}},
    \end{aligned}
\end{equation}
where~$V=v_h/v_\mathrm{dm}$ and~$\displaystyle \Gamma[n,x]=\int_x^\infty dt e^{-t} t^{n-1}$ is the incomplete Gamma function. 
Interestingly, in equation~\eqref{eq:late_time_bfr_per_halo}, the dependence on the compact objects mass factorizes out of the merger rate, i.e., as pointed out in ref.~\cite{scelfo:gwxlssI}, we can rescale the binary formation rate for different PBH mass distributions and abundances just by recomputing the quantity
\begin{equation}
    \mathcal{M}_\mathrm{PBH} = \int dM_1 dM_2 \frac{d\Phi_\mathrm{PBH}}{dM_1}\frac{d\Phi_\mathrm{PBH}}{dM_2} \left(\frac{M}{\mu}\right)^{5/7},
\end{equation}
where a useful reference value is given in the MMD case by~$\mathcal{M}^\mathrm{MMD}_\mathrm{PBH}=4^{5/7}$, independently from where the mass distribution is peaked. 

We follow ref.~\cite{cholis:orbitaleccentricity} to model the initial properties of the binaries: they form with an initial semi-major axis and eccentricity
\begin{equation}
    a_\mathrm{ini} = -\frac{G M\mu}{2E_{\rm ini}}, \qquad e_\mathrm{ini} = \sqrt{1+2\frac{E_{\rm ini} b^2 v_{\rm rel}^2}{G^2 M^2\mu}},
\end{equation}
where the energy of the binary at the beginning of the inspiral phase
\begin{equation}
    E_{\rm ini} = \frac{\mu v_{\rm rel}^2}{2}-\frac{85\pi G^{7/2}}{12\sqrt{2}c^5}\frac{\mu^2 M^{5/2}}{r_{\rm p}^{7/2}}
\end{equation}
depends on the distance of closest approach
\begin{equation}
    r_{\rm p} \simeq \frac{b^2 v_{\rm rel}^2}{2GM}\left(1-\frac{b^2v_{\rm rel}^4}{4G^2M^2}\right),
\end{equation}
and on the impact parameter~$b$. The impact parameter is assumed to be uniformly distributed between the minimum and maximum impact parameters
\begin{equation}
    b_\mathrm{min} = \sqrt{12} \frac{GM}{c v_\mathrm{rel}}, \qquad b_\mathrm{max} = \left(\frac{340\pi}{3}\right)^{1/7} \frac{GM^{6/7}\mu^{1/7}}{c^2} \left(\frac{v_\mathrm{rel}}{c}\right)^{-9/7}.
\end{equation}
Finally, the time delay between binary formation and merge is given by
\begin{equation}
    t_d = \frac{3}{85}\frac{a_\mathrm{ini}^4c^5}{G^3 M^2 \mu} \left(1-e_\mathrm{ini}^2\right)^{7/2},
\end{equation}
therefore, by statistically sampling the initial properties of the binaries in a similar fashion to what presented in ref.~\cite{bellomo:classgwb}, we can determine the time delay probability distribution function~$p(t_d)$ entering in equation~\eqref{eq:late_time_merger_rate_density}.

\begin{figure}[t]
\centerline{
\includegraphics[width=\columnwidth]{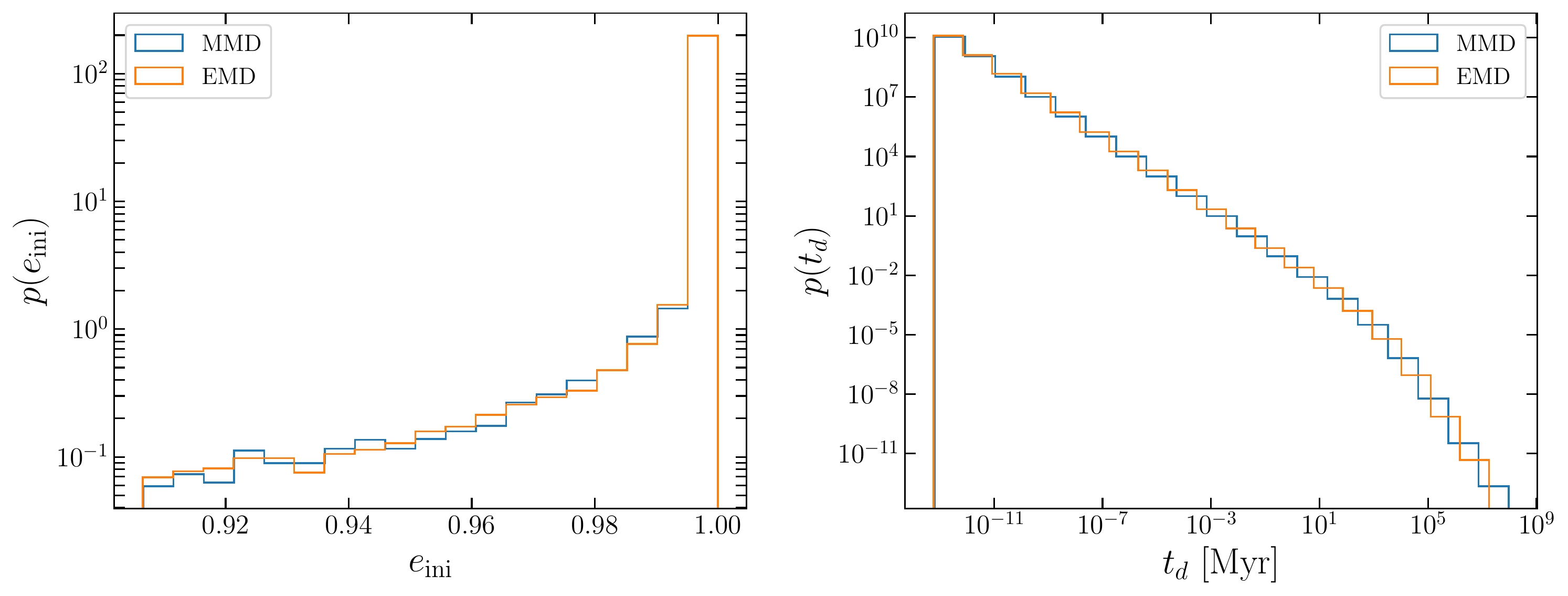}}
\caption{Initial eccentricity (\textit{left panel}) and time delay (\textit{right panel}) probability distribution function for our choice of benchmark MMD (\textit{blue} lines) and EMD (\textit{orange} lines).}
\label{fig:lpbh_properties}
\end{figure}

We show in figure~\ref{fig:lpbh_properties} the initial eccentricity and time delay probability distribution function obtained for our choice of MMD and EMD as example of our study of the binary properties.
First of all we note by comparing the MMD and EMD histograms that binaries properties are quite independent from the details of the PBH mass function.
As expected all the binaries are created with very high eccentricities.
Second, we prove that the time delay probability distribution function scales approximately as~$p(t_d) \propto t^{-1}_d$, with minimum delay of~$t_{d,\mathrm{min}} \simeq 1\ \mathrm{s}$ and maximum one that can exceed the lifetime of the Universe, corresponding to binaries that never merge.
Therefore, we conclude that the maximum time delay should be set to be~$t_{d,\mathrm{max}} = t(z)$, with~$t(z)$ being the age of the Universe at redshift~$z$.


\section{Optimistic cross-correlation constraints}
\label{app:optimistic_mg_constraints}

\begin{table}[h]
    \centering
    \begin{tabular}{|c|c|c|c|c|}
    \hline
    Tracers & $k_\mathrm{mg}\ [\mathrm{Mpc}^{-1}]$ & $\sigma_{\mu_0}$ & $\sigma_{\eta_0}$ & $\sigma_{b^\mathrm{eff}_\mathrm{GW}}$ \\
    \hline
    \hline
    \multirow{3}{*}{ET2CE} & $10^{-1}$ & $0.39\ (0.30)$ & $2.80\ (1.54)$ & $1.54\ (0.55)$ \\
                           & $10^{-2}$ & $1.00\ (0.79)$ & $3.06\ (2.16)$ & $1.37\ (0.13)$ \\
                           & $10^{-3}$ & $28.68\ (28.52)$ & $37.10\ (36.82)$ & $1.01\ (0.09)$ \\
    \hline
    \multirow{3}{*}{SW$\times$ET2CE} & $10^{-1}$ & $0.02\ (0.02)$ & $0.15\ (0.06)$ & $0.10\ (0.04)$ \\
                                     & $10^{-2}$ & $0.11\ (0.10)$ & $0.28\ (0.25)$ & $0.08\ (0.04)$ \\
                                     & $10^{-3}$ & $2.91\ (2.23)$ & $2.23\ (2.90)$ & $0.07\ (0.04)$ \\
    \hline
    \multirow{3}{*}{DL$\times$ET2CE} & $10^{-1}$ & $0.02\ (0.02)$ & $0.10\ (0.06)$ & $0.07\ (0.05)$ \\
                                     & $10^{-2}$ & $0.11\ (0.11)$ & $0.29\ (0.24)$ & $0.06\ (0.05)$ \\
                                     & $10^{-3}$ & $3.51\ (3.51)$ & $2.93\ (2.92)$ & $0.06\ (0.05)$ \\
    \hline
    \end{tabular}\\
    \vspace{0.3cm}
    \begin{tabular}{|c|c|c|}
    \hline
    Tracers & $\sigma_{\Omega_\mathrm{rc}}$ & $\sigma_{b^\mathrm{eff}_\mathrm{GW}}$ \\
    \hline
    \hline
    ET2CE           & $2.18\ (1.56)$ & $0.98\ (0.34)$ \\
    SW$\times$ET2CE & $0.11\ (0.03)$ & $0.07\ (0.04)$ \\
    DL$\times$ET2CE & $0.10\ (0.04)$ & $0.06\ (0.05)$ \\
    \hline
    \end{tabular}
\caption{Marginalized errors on Modified Gravity (\textit{upper table}) and nDGP (\textit{lower table}) parameters for single (GW only) and multi-tracer analysis. 
SW and DL refer to ``shallow and wide'' and ``deep and localized'' galaxy surveys, respectively.
Errors refer to our optimistic benchmark model. 
Numbers in parenthesis are the marginalized errors obtained assuming perfect knowledge of the standard cosmological parameters.}
\label{tab:optimistic_mg_constraints}
\end{table}

Table~\ref{tab:optimistic_mg_constraints} shows the marginalized errors on Modified Gravity and nDGP parameters with and without considering perfect knowledge of the standard cosmological parameters and assuming an optimistic version of our benchmark model.
The optimistic model assumes that the number of observed GW events per year is twice that of the conservative case (hence the shot noise is roughly half of that of the conservative case), in line with current estimates of certain astrophysical BH models~\cite{iacovelli:gw3genforecast}.


\section{Mixed black hole population for second generation detectors}
\label{app:mixed_BBH_2G}

\begin{figure}[h]
    \centerline{
    \includegraphics[width=\columnwidth]{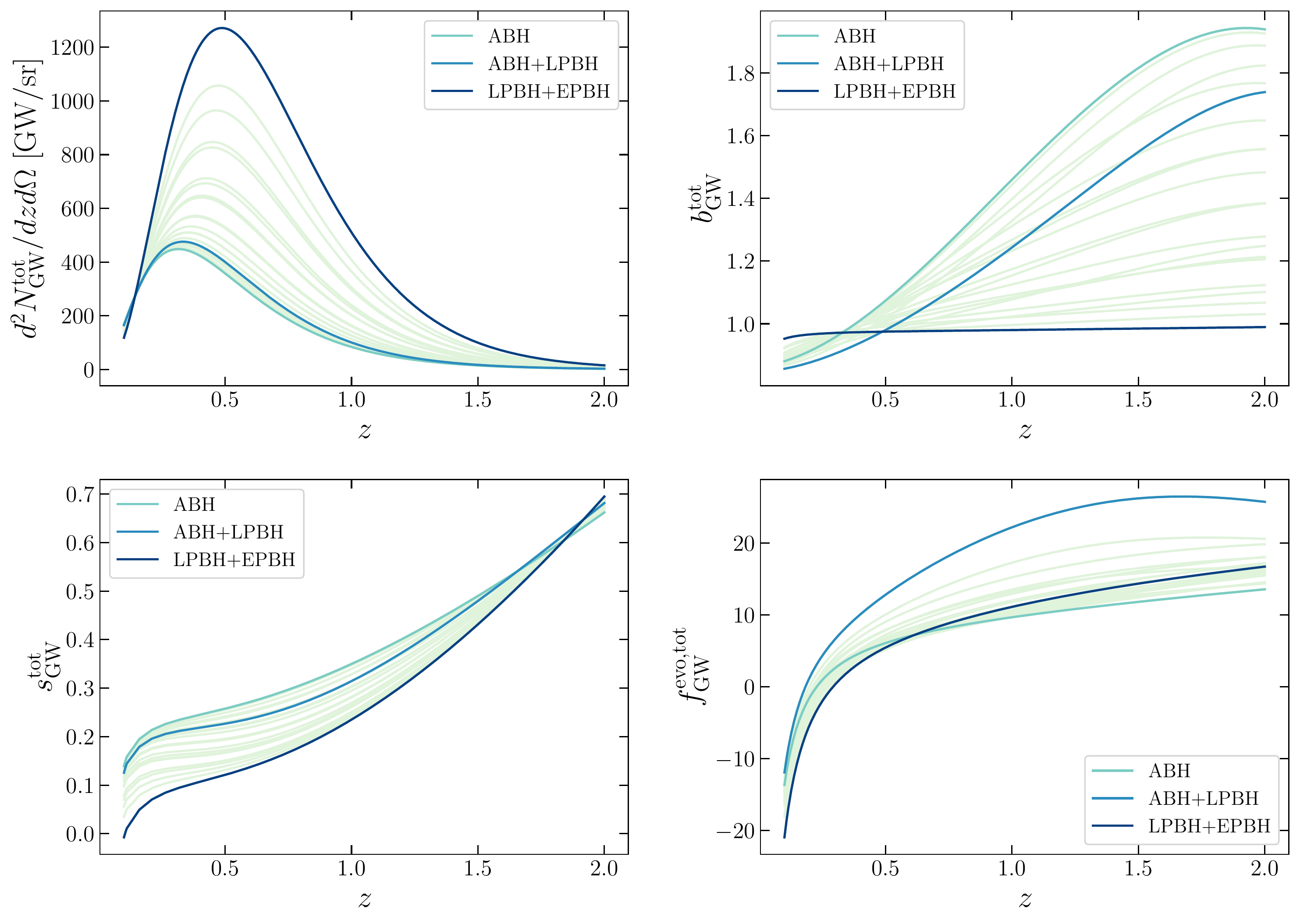}}
\caption{GW mixed population redshift distribution (\textit{top left panel}), bias (\textit{top right panel}), magnification bias (\textit{bottom left panel}) and evolution bias (\textit{bottom right panel}) for different values of~$f_\mathrm{PBH}$ and~$\mathcal{A}_m$.
In this plot we consider only events detected by a second generation GW detector network.
The \textit{light blue}, \textit{blue} and \textit{dark blue} lines correspond to the astrophysical BHs only ($f_\mathrm{PBH}=\mathcal{A}_\mathrm{m}=0$), mixed astrophysical and late time PBH ($f_\mathrm{PBH}=1,\ \mathcal{A}_\mathrm{m}=0$), and PBHs only ($f_\mathrm{PBH}=1,\ \mathcal{A}_\mathrm{m}=18$) scenarios, respectively.
The \textit{green} lines refer to mixed models where different populations contribute to the total number of detected events.} 
\label{fig:BH_2G_mixed_models}
\end{figure}

Figure~\ref{fig:BH_2G_mixed_models} shows the total redshift distribution, bias, magnification bias and evolution bias for different models spanning the~$f_\mathrm{PBH}-\mathcal{A}_\mathrm{m}$ parameter space for a second generation detector network, along with three reference scenarios corresponding to the astrophysical BHs only ($f_\mathrm{PBH}=\mathcal{A}_\mathrm{m}=0$), mixed astrophysical and late time PBH ($f_\mathrm{PBH}=1,\ \mathcal{A}_\mathrm{m}=0$), and PBHs only ($f_\mathrm{PBH}=1,\ \mathcal{A}_\mathrm{m}=18$) cases.
In all the scenarios we observe a redshift distribution peaked~$z\simeq 0.3-0.5$, with differences between models suppressed by the detector volume selection effects.
The dispersion of bias, magnification bias and evolution bias values around the peak of the redshift distribution is rather small, lowering the possibility of disentangling how much each component contributes to the total population of observed events.


\bibliography{bibliography}
\bibliographystyle{utcaps}

\end{document}